\documentclass[intlimits,twoside,a4paper]{article}

\usepackage[cp1251]{inputenc}

\usepackage{cmpj3}

\issue{2018}{21}{1}{13002}
\doinumber{10.5488/CMP.21.13002}

\title[Phase field modelling voids nucleation and growth in binary systems]%
{Phase field modelling voids nucleation and growth in binary systems}%

\author[D.O. Kharchenko \textit{et al.}]{D.O.~Kharchenko\refaddr{label1}, V.O.~Kharchenko\refaddr{label1}, Y.M.~Ovcharenko\refaddr{label1}, O.B.~Lysenko\refaddr{label1}, I.A.~Shuda\refaddr{label2}, L. Wu\refaddr{label3}, 
R. Pan\refaddr{label3}}
\addresses{\addr{label1} Institute of Applied Physics of the
National Academy of Sciences of Ukraine,\\
58 Petropavlivska St., 40000 Sumy, Ukraine
\addr{label2} Sumy State University, 2 Rimskii-Korsakov St., 40007
Sumy, Ukraine
\addr{label3} The First Institute, Nuclear Power Institute of China, 328, the 1st
Section, Changshundadao Road, Shuangliu, Chengdu, China}

\date{Received December 18, 2017}

\begin{document}

\maketitle

\begin{abstract}
We present a comprehensive study of voids formation, nucleation and growth in a prototype model of binary alloys subjected to irradiation  by using a combined approach based on phase field and rate theories. It is shown that voids formation is caused by interaction of irradiation-produced vacancies through elastic deformation of a lattice and vacancy coupling with composition field of the alloy. Phase diagrams illustrating the formation of states related to  solid solution, phase decomposition, and patterning are obtained. 
Formation of voids from supersaturated ensemble of vacancies is accompanied by composition rearrangement  of alloy components.
It was found that elastic inhomogeneity leading to the formation of anisotropic precipitates in an initially prepared binary alloy results in the formation of a void super-lattice under irradiation.
It was shown that voids nucleate and grow with dose according  to diffusion controlled precipitation processes, where universal dynamics of voids growth is revealed. 
Estimations of main quantitative and statistical characteristics of voids 
by using material parameters relevant to most of alloys and steels give good agreement with experimental observations.

\keywords binary alloy, patterning, irradiation, point defects, voids
\pacs 05.10.-a, 89.75.Kd, 61.80.Lj, 61.72.J-, 61.72.Qq
\vspace{-1mm}
\end{abstract}

\section{Introduction}
\vspace{-1mm}
An evolution of defect microstructure accompanied by voids formation and growth in structural alloys is a complex process responsible for  the long-term performance of such alloys. It is well known that the mechanisms governing void nucleation relate to excess of vacancies and interstitials produced in collision cascades \cite{ZS1999}. Nonequilibrium point defects are capable of recombinating, annihilating at sinks, segregating into clusters \cite{clusters1,clusters2} which results in the formation of dislocation loops, defect walls \cite{walls}, and voids \cite{DM2003}. Displacement cascades occurring during several picoseconds increase  point defect concentrations up to several orders comparing to their thermal (equilibrium) values.  Combined effect of diffusion and interaction of supersaturated ensemble of mobile (point) defects leads to self-organization processes  with formation of defects of high dimensions (stacking faults, voids, defect walls) and microstructure transformation of the  crystalline systems studied \cite{Walgraef1,Was}. Voids can be organized into periodic arrays by forming a void super-lattice. This phenomenon was discovered by Evans \cite{Evans}. 
Partially ordered void super-lattices have  been reported
for irradiated pure materials, for example, Mo \cite{Evans}, Al \cite{voidsAl}, Nb \cite{voidsNb},  Ni \cite{voidsNi},  and  alloys: Ni-Al \cite{voidsNiAl}, Cu-10\%Ni \cite{voidsCuNi}, Nb-Zr \cite{voidsNbZr}, 
Mo-0.5Ti \cite{voidsMoTi}, 
stainless steel \cite{voids_exp1,voidsSS}  irradiated to damage levels of 
$10{-}80$~dpa (displacements per atom) \cite{GWZ}. It was found that typical void size is $2{-}7$~nm in pure materials with void super-lattice parameter  $20{-}30$~nm, whereas for steels, the void size is  around $30{-}100$~nm, depending on irradiation conditions (dose rate and temperature). Experimental studies have shown that typical void concentration is  $10^{15}{-}10^{16}$~cm$^{-3}$ \cite{Was,GWZ}.

As usual, voids  evolution is considered as three separate processes reduced to nucleation, growth, and coarsening. In the framework of the standard nucleation theory, only spatially averaged point defect concentrations are used in the analysis \cite{R1971,W1972,BBH1976,MB1980,WS1982}. Such approaches are based on mean field reaction rate theory by neglecting spatial rearrangement of point defects. Therefore, it cannot provide information on the microstructure evolution in irradiated alloys. The emergence of gradients of external factors (stress, temperature, etc.) induces migration of voids, formation of void lattices in systems  under irradiation. Therefore, the usage of models of spatially extended systems of point defects can give more insight onto the dynamics of self-organization of defects in such materials \cite{GW1993,Walgraef1}.

A progress in the studies of extended systems of point defects in the framework of continuum approach based on reaction rate theory  was achieved  during a few recent decades (see, for example, reviews \cite{Walgraef1,GWZ,Golubov} and citations therein). In \cite{Walgraef96} it was shown that by considering a spatially extended system of defects one can monitor the dynamics of dislocation loop densities. Point defects patterning under different irradiation conditions were presented in \cite{EPJB2012,CMPh2013,UJP2013,REDS2014}, where the results of numerical simulations of self-organization of point defects into nanometer clusters and defect walls  are discussed. Spatially extended models of the system of point defects were used to study voids formation \cite{EPJB2016} and grains fragmentation under sustained irradiation \cite{PRE2014}.  Concentration of point defects and  loop densities were considered as the main modes in the above approaches. 

Nowadays, a very ambitious idea of studying the radiation effects  is a self-consistent combination of different methods working correctly on specific time and length limits. Such multiscale modelling is a quite complicated problem  requiring a huge amount of computation  resources and  specific information of targets (alloys), for instance, energetic properties of defects and impurities, interatomic potentials, elastic moduli, phase diagrams, etc.
Some hybrid approaches can be used to study fundamental  mechanisms governing the microstructure evolution  by combining several methods working correctly at least on two time and length scales. Among them one can point out a phase field method based on Allen-Cahn and Cahn-Hilliard approaches \cite{Cahn61,Cahn63,Cahn58}
allowing one to  monitor phase transitions \cite{LiuBellon2002}, precipitation \cite{EB2001}, patterning \cite{EB2000}, voids formation \cite{YuLu2005,Rokkam2009,VoidsReview2017}, as well as a  change in mechanical properties at irradiation \cite{REDS2015}.

Voids and gas bubbles formation and their evolution can be studied by combining spatially-resolved reaction rate theory and phase field method  (see \cite{VoidsReview2017} and citations therein). The phase field method  correctly predicts the relationships between
microstructures and properties of the studied systems. In the framework of this approach, the  void formation processes were described by introducing the phase field  as an  order parameter to separate the void  phase from the matrix phase. A generalization of this theory onto problems of voids formation in polycrystalline metals was reported in \cite{Part1}. A further development was provided in \cite{Part2}, where gas bubbles formation were studied. A phase field model for void lattice formation under irradiation was proposed in \cite{HH2009}. Multi-component systems with voids formation are usually studied by molecular dynamics and Monte-Carlo simulations. In \cite{BBA2015},  a mixed discrete-continuous phase field approach was proposed to capture the stochastic
character of the production of a defect cluster in displacement cascades in two- and multi-component alloys. The development of Darken's approach in binary systems under irradiation, where vacancies are capable of organizing into voids  was discussed in \cite{Gusak}. Properties of patterning in a two-component system caused by vacancy rearrangement at irradiation and the corresponding discussions of the Kirkendall effect  were provided in \cite{UJP2016,REDS2016}.  In most of the works related to the study of the void evolution, point defects were described by their populations and all classifications of defect patterns were done by considering  the corresponding values of point defect concentrations, whereas no special quantity identifying a different kind of objects having dimensions larger than zero was introduced. Therefore, a problem related to identification and studying the evolution of voids as separated objects in a system of point defects  remains actual. At the same time, a problem to describe a composition change in irradiated alloys with voids formation is still open.

In this paper we study the defect structure evolution during nucleation and growth of voids in a binary alloy system where it will be shown that voids formation is accompanied by the local change in the composition of the alloy studied. We generalize a standard phase field approach by taking into account the dynamics of solute concentration and population of point defects produced at irradiation in collision cascades. In our approach we combine the phase field method and reaction rate theory to take into account the production of defects, their recombination and annihilation. We consider the case when vacancies are capable of deforming the crystalline lattice due to their interaction with elastic fields. 
A novelty of this work is in a development of  the phase field approach for  binary systems subjected to irradiation. The derived approach allows one to monitor the voids formation due to rearrangement of a supersaturated ensemble of vacancies accompanied by compositional patterning. In our consideration we describe the structural disorder formation leading to athermal atomic mixing by considering the dynamics of point defects  continuously produced during irradiation. In such a case, athermal atomic mixing  emerges as a process accompanied by defects  production. By using this physical mechanism of athermal mixing we compute phase diagrams illustrating transitions between the states related to solid solution, phase decomposition and patterning of a binary alloy at different irradiation conditions characterized by temperature and damage rate. Here  the patterns formation corresponds to the formation of voids. It will be shown that the obtained phase diagrams relate well to the results calculated by using an assumption of ballistic diffusion in the phase field models with a help of the Cahn-Hilliard theory (see \cite{Martin,Vaks,EB2000,EB2001,BBA2015}).
In our computations we initially prepare the binary alloys as a target with equilibrium vacancy concentration by assuming that all interstitials are relaxed at sinks. Next, we consider this target under  sustained irradiation to study the voids formation, their  nucleation, and growth. 

The work is organized in the following manner. In section~\ref{2}, we derive a generalized phase field model for the dynamics of defect populations connected with  solute concentration, lattice deformation and phase field. 
Linear stability analysis is given in section~\ref{3}, where there is obtained a phase diagram illustrating the intervals for irradiation temperature and damage rate related to the formation of solid solution with stochastic vacancy distribution, phase decomposition with segregation of nonequilibrium vacancies at interfaces and patterning of the solid solution due to the voids formation. 
In section~\ref{4}, the results of numerical tests are presented. Here we provide a statistical analysis of the voids nucleation and growth. Qualitative estimations of the obtained numerical results with experimental data are given.  We conclude in section~\ref{5}.

\section{Model} \label{2}

By considering a binary system with atoms of sorts A and B subjected to sustained irradiation, one deals with the corresponding  concentrations $c_{\text A}$ and $c_{\text B}$, and a concentration of point defects. Next, we introduce vacancy concentration  $c_{\text v}$ and  two types of the corresponding interstitials with concentrations $c_{\text i}^{\text A}$  and $c_{\text i}^{\text B}$ (dumbbells are not considered here). In this study we consider  network dislocations, vacancy and interstitial loops as sinks of point defects. Network dislocation density remains constant during irradiation, whereas vacancy loop densities evolve in time.

By using the reaction rate theory  one can consider the dynamics of each kind of defects by taking into account the defect production in collision cascades, annihilation at sinks and mutual recombination \cite{GWZ,Was}. 
The dynamics of interstitial concentrations is described by an equation of the form:
\begin{equation}\label{evol_ci}
\begin{split}
\partial_t c_{\text i}^{\text{A,B}}&=\mathcal{K}_0(1-\varepsilon_{\text i})c_{\text{A,B}}-D_{\text i}\rho_{\text N}\mu(\rho_{\text{i,v}})c_{\text i}^{\text{A,B}} -\alpha_{\text r}(c_{\text v}-c_{\text{v}0})c_{\text i}^{\text{A,B}}-\nabla\cdot \mathbf{J}^{\text{A,B}}_\text{i}. 
\end{split}
\end{equation} 
The first term is responsible for the production of interstitials from the solute, the second one relates to annihilation of the defects with  sinks, the third one corresponds to recombination. 
Here, $\mathcal{K}_0$ is the defect production rate measured in dpa/s (displacement per atom per second),  $D_{\text i}$ is the interstitial diffusivity,  $\rho_{\text N}$ is the network dislocation density, $\mu(\rho_{\text{i,v}})=1+{\rho_{\text i}}/{\rho_{\text N}}+{\rho_{\text v}}/{\rho_{\text N}}$ is the renormalized sinks strength defined through interstitial and vacancy loop densities $\rho_{\text i}$ and $\rho_{\text v}$, respectively,   $\alpha_{\text r}=4\piup r_0D_{\text i}/\Omega_0$ is the recombination rate expressed through the capture radius $r_0\simeq (3{-}5)a$ and the atomic volume $\Omega_0$, $a$ is the lattice parameter,  $c_{\text{v}0}\simeq \exp(-E_{\text v}^{\text f}/T)$ is the equilibrium vacancy concentration,  $E_{\text v}^{\text f}$ and $T$ are vacancy formation energy and temperature measured in energetic units,   $\mathbf{J}_{\text i}^{\text{A,B}}$ is the  flux of the corresponding interstitials.  

Evolution equation for the vacancy concentration  is of the form 
\begin{equation}\label{evol_cv}
\begin{split}
\partial_t c_{\text v}&=\mathcal{K}_0(1-\varepsilon_{\text v})-\left[D_{\text v}\rho_{\text N}\mu(\rho_{\text{i,v}})+\alpha_{\text r}(c_{\text i}^{\text A}+c_{\text i}^{\text B})\right](c_{\text v}-c_{\text{v}0})-\nabla\cdot \mathbf{J}_{\text v}.
\end{split}
\end{equation}
The first three terms correspond to the formation of vacancies, annihilation and recombination,  respectively;  $\mathbf{J}_{\text v}$ is the  flux of vacancies;  $\varepsilon_{\text{i,v}}$ in equations~(\ref{evol_ci}), (\ref{evol_cv}) relates to cascade collapse efficiencies. 

We consider that interstitials are produced when atoms leave their positions in lattice. Therefore, this effect should be incorporated into the evolution equation for concentration of atoms of each sort. It takes  the form 
\begin{equation}\label{evol_sol}
\begin{split}
\partial_tc^{\text{A,B}}=&-\mathcal{K}_0(1-\varepsilon_{\text i})c^{\text{A,B}}+[D_{\text i}\rho_{\text N}\mu(\rho_{\text{i,v}})+\alpha_{\text r} (c_{\text v}-c_{\text{v}0})]c_{\text i}^{\text{A,B}}-\nabla\cdot \mathbf{J}_{\text{A,B}}.
\end{split}
\end{equation}
Here, the first term relates to the production of interstitials of one sort (an atom knocked in cascades becomes interstitial), the second term relates to annihilation of interstitials with sinks (the first term in square brackets) and  recombination when a single interstitial of one sort recombines with the vacancy becoming an atom of the corresponding sort in the solute (the second term in square brackets), 
$\mathbf{J}_{\text{A,B}}$ relates to a flux of the corresponding component. 

The dynamics  of  interstitial and vacancy loop
densities is described by the model proposed in \cite{BEK}. In our case it takes the form:
\begin{equation}\label{evol_sinks}
\begin{split}
 &\partial_{t}\rho_{\text i}=\frac{1}{r_{\text i}|\mathbf{b}|} \left\{ \mathcal{K}_0\varepsilon_{\text i}-\rho_{\text i}\left[D_{\text v}(c_{\text v}-c_{\text{vL}})-D_{\text i}\big(c_{\text i}^{\text A}+c_{\text i}^{\text B}\big)\right]\right\};\\
 &\partial_{t}\rho_{\text v}=\frac{1}{ r_{\text v}|\mathbf{b}|}\left\{\mathcal{K}_0\varepsilon_{\text v}-\rho_{\text v}\left[D_{\text i}\big(c_{\text i}^{\text A}+c_{\text i}^{\text B}\big)-D_{\text v}(c_{\text v}-c_{\text{vL}})\right]\right\}.
\end{split}
\end{equation}
Here, $\mathbf{b}$ is the Burgers vector, $r_{\text{i,v}}$  denotes 
the initial interstitial/vacancy loop radius.
The line density of loops $\rho_{\text{i,v}}$ is given by the mean loop radius, $r_{\text{i,v}}$, and the number density of loops, $N_{\text{i,v}}$, as follows $\rho_{\text{i,v}}=2\piup r_{\text{i,v}}N_{\text{i,v}}$. As was discussed in \cite{SemWoo}, the mean  vacancy loop radius $r_{\text v}$ is around  half the initial loop radius, i.e.,  $r_{\text v}\simeq r_{0\text{v}}/2$, where $r_{0\text{v}}=(n_{\text v}\Omega/\piup|\mathbf{b}|)^{1/2}$, $n_{\text v}\simeq 30{-}50$ is the number of vacancies in the loop at creation. For the interstitial loop radius we use the relation: $r_{\text i}=[r_{\text i}(n_{\text i})+r_{\text i}(n_{0\text{i}})]/2$, where the number of atoms in interstitial clusters at the moment of creation is  $n_{\text i}\approx 6{-}10$, $n_{0\text{i}}\approx 4{-}5$ is the critical number of interstitials in clusters treated as small loops. Below $n_{0\text{i}}$, clusters become  mobile as single interstitials \cite{ERS,OBSSG}. The quantity $c_{\text{vL}}=c_{\text{v}0}\exp[(\gamma_{\text{SF}}+F_{\text e})\Omega_0/|\mathbf{b}|T]$  denotes the equilibrium vacancy concentration near the vacancy dislocation loops, where $\gamma_{\text{SF}}$ is the stacking fault energy, $F_{\text e}=\mu_{20}\Omega_0\ln(r_{\text L}/|\mathbf{b}|+1)/[4\piup(1-\nu)|\mathbf{b}|(r_{\text L}+|\mathbf{b}|)]$ is the elastic line energy, $\nu$ is the Poisson ratio, $r_{\text L}=r_{\text{v}0}/2$.   These line defects are assumed immobile.

Interstitials produced in cascades move to sinks faster than vacancies due to their diffusivities difference $D_{\text i}/D_{\text v}\gg1$. Moreover, at high temperatures, the concentration of interstitials attains stationary value faster than the vacancy concentration (see discussions in \cite{Was}). Another reason responsible for the difference in time scales for vacancy and interstitial concentrations is the difference in cascade collapse efficiencies, $\varepsilon_{\text i}$ and $\varepsilon_{\text v}$, where one usually uses $\varepsilon_{\text i}/\varepsilon_{\text v}>1$. It means that effective generation rates for non-clustered vacancies and interstitials are different and they cannot reach the steady state at the same time.   Therefore,  by considering the high temperature limit and the difference in cascade collapse efficiencies, next, we adiabatically eliminate the interstitial concentration by putting $(D_{\text v}\rho_{\text N})^{-1}\partial_t c_{\text i}^{\text{A,B}}\simeq0$. It allows us to use an algebraic relation for the concentration of interstitials.
After some algebra we obtain a total system of dynamical equations governing the evolution of the considered fields under sustained irradiation  in the form:
\begin{equation}
\begin{split}
&\partial_tc^{\text{A,B}}=-\nabla\cdot \mathbf{J}_{\text{A,B}};\\
&\partial_t c_{\text v}=\mathcal{K}(1-\varepsilon_{\text v})-\mu(\rho_{\text{i,v}})(c_{\text v}-c_{\text{v}0})-\frac{\mathcal{K}(1-\varepsilon_{\text i})(c_{\text v}-c_{\text{v}0})(1-c_{\text v})} {\delta\mu(\rho_{\text{i,v}})+(c_{\text v}-c_{\text{v}0})}-\nabla\cdot \mathbf{J}_{\text v};\\
&  \tau_{\rho_{\text i}}\partial_{t}\frac{\rho_{\text i}}{{\rho_{\text N}}}= \mathcal{K}\varepsilon_{\text i}-\frac{\rho_{\text i}}{{\rho_{\text N}}}\left[\frac{\mathcal{K}(1-\varepsilon_{\text i})(1-c_{\text v})}{\mu(\rho_{\text{i,v}})+\delta^{-1}(c_{\text v}-c_{\text{v}0})}+(c_{\text v}-c_{\text{vL}})\right];\\
& \tau_{\rho_{\text v}}\partial_{t}\frac{\rho_{\text v}}{\rho_{\text N}}=\mathcal{K}\varepsilon_{\text v}-\frac{\rho_{\text v}}{\rho_{\text N}}\left[\frac{\mathcal{K}(1-\varepsilon_{\text i})(1-c_{\text v})}{\mu(\rho_{\text{i,v}})+\delta^{-1}(c_{\text v}-c_{\text{v}0})}-(c_{\text v}-c_{\text{v}0})\right].
\end{split}
\end{equation} 
Here, we use a dimensionless dose rate $
\mathcal{K}\equiv\mathcal{K}_0/D_{\text v}\rho_{\text N}$. The physical time is measured in units  $(D_{\text v}\rho_{\text N})^{-1}$.  
We consider  $\tau_{\rho_{\text i}}\equiv{r_{\text i}|\mathbf{b}|\rho_{\text N}}$, and $\tau_{\rho_{\text v}}\equiv{r_{\text v}|\mathbf{b}|\rho_{\text N}}$ as time scales for the evolution of sink densities.  The quantity $\delta\equiv\rho_{\text N}\Omega_0/4\piup r_0$ at fixed  material parameters is considered as a small parameter, i.e., $\delta\ll 1$. 

Next, let us consider the free energy of a system with vacancies and define the corresponding diffusion fluxes. By introducing a composition difference $\psi\equiv c_{\text A}-c_{\text B}$ with restriction $c_{\text B}+c_{\text A}+c_{\text v}=1$, 
we use a total free energy of unirradiated sample  $F_0=\int_V f(\psi, c_{\text v}, \mathbf{u})\rd\mathbf{r}$
depending on  the composition difference
$\psi=\psi(\mathbf{r})$, vacancy concentration $c_{\text v}(\mathbf{r})$,  and the elastic displacements vector
$\mathbf{u}=(u_x,u_y)$. The total free energy density $f$ takes into account a
component related to the chemical free energy and a corresponding elastic
contribution caused by the elastic mismatch via the Vegard law. Following
\cite{OnukiBook2002,OnukiPhysRevE68,OnukiPhysRevB70,OnukiPhysRevB72}, the
free energy density $f$ reads:
\begin{equation}\label{eq_f}
	f=\frac{1}{\Omega_0} f(\psi,c_{\text v})
  +f_{\text{el}}(\mathbf{u},\psi,c_{\text v}).
\end{equation}

By using Bragg-Williams approach, a chemical free energy density $f(\psi,c_{\text v})$ acquires a form  
\begin{align}\label{f_dens1}
f(\psi,c_{\text v})&=E_{\text v}^{\text f} c_{\text v}-\frac{Z\omega_0}{2}\psi^2\nonumber\\
           &+T\left[\frac{1}{2}(1+\psi-c_{\text v})\ln (1+\psi-c_{\text v})+\frac{1}{2}(1-\psi-c_{\text v})\ln (1-\psi-c_{\text v})+c_{\text v}\ln c_{\text v}\right].
\end{align}
Here,  $E_{\text v}^{\text f}$ is the vacancy formation energy, $Z$ is the coordination number, the ordering energy  $\omega_0=2V_{\text{AB}}-V_{\text{AA}}-V_{\text{BB}}$ is expressed through interaction energies $V_{\text{AA}}$, $V_{\text{BB}}$, and $V_{\text{AB}}$.

The  term $f_{\text{el}}(\mathbf{u},\psi,c_{\text v})$ is responsible for the connection between  composition, vacancy concentration and true elastic contribution. It can be decomposed as  
\begin{equation}
f_{\text{el}}(\mathbf{u},\psi,c_{\text v})=f^{(0)}_{\text{el}}(\mathbf{u},\psi,c_{\text v})+f^{(1)}_{\text{el}}(\mathbf{u},\psi).
\end{equation}  
The first term  takes into account the feedback between deformation,  composition and vacancy concentration, the second one corresponds to the elastic energy change under the assumption that the elastic moduli depend on the composition field. 
For the first part we have 
\begin{equation}\label{eq_f_a}
f^{(0)}_{\text{el}}(\mathbf{u},\psi,c_{\text v})=\alpha e_1\psi+{\beta}e_1c_{\text v}.
\end{equation}
Here, the term with  the
constant $\alpha>0$ describes feedback of the composition difference $\psi$ and
the elastic displacements in the dilation strain $e_1=\nabla \cdot \mathbf{u}$.
This term can be explained by compositional inhomogeneity with the formation of
the Cottrell atmosphere 
in the one-phase state and incoherent inclusions such
as $\gamma'$-precipitates in the $L1_2$ structure in the two-phase state
\cite{OnukiPhysRevB70}. As was shown in \cite{OnukiBook2002}, this term
arises due to generalization of  Cahn and Khachaturyan theories \cite{Khachaturyan} admitting
that bulk and shear moduli are constants. In the generalized case, one assumes
that a bulk modulus is a constant, whereas a shear modulus is a linear function
of the composition field\footnote{Technical details illustrating derivation of
the free energy density with the coupling term can be found in
\cite{OnukiBook2002}.}.  The  term with $\beta<0$ in equation~(\ref{eq_f_a}) relates to deformations of an elastic continuum caused by defects, where $\beta=K\Omega_{\text v}$, $\Omega_{\text v}<0$ is the dilatation volume for vacancies, $K$ is the bulk modulus \cite{Mirzoev}. 
For the component $f^{(1)}_{\text{el}}(\mathbf{u},\psi)$, we admit 
\begin{equation}\label{eq_f_el}
 f^{(1)}_{\text{el}}(\mathbf{u},\psi)=\frac 1 2 K e_{1}^2+\Phi(\psi,e_2,e_3),
\end{equation}
where the first term with the bulk modulus $K$ determines the dilation elastic
energy, the second term 
\begin{equation}\label{eq_f_el1}
\Phi(\psi,e_2,e_3)=\frac{\mu_2}{4\piup^2}\left[1-\cos (2\piup e_2)\right] +\frac{\mu_3}{4\piup^2}\left[1-\cos (2\piup e_3)\right]
\end{equation}
relates to anisotropic deformations with the tensile modulus $\mu_2$ and the
shear modulus $\mu_3$, respectively, and defines the shear elastic energy. 
In the two-dimensional case considered below,  $\Phi$ depends on 
tetragonal strain $e_2=\partial_x u_x-
\partial_y u_y$ and shear strain $e_3=\partial_x u_y+\partial_y u_x$ expressed
through the elastic displacements~$\mathbf{u}$ according to the Cauchy relations,
as usual. 
 According
to \cite{OnukiBook2002}, we assume that elastic moduli $\mu_2$ and $\mu_3$
are the composition dependent functions, i.e., $\mu_k=\mu_{k0}+\mu_{k1}\psi$, $k=2,3$,
where $\mu_{k0}$ and $\mu_{k1}$ are constants. In terms of usual elastic moduli, one has: $\mu_{20}=\frac{1}{2}(C_{11}-C_{12})$, $\mu_{30}=C_{44}$. Following
\cite{OnukiPhysRevB70} if $\mu_{21}, \mu_{31}> 0$, then the regions with
large $\psi$ are harder (hard phase), whereas regions with small $\psi$ are
softer (soft phase). Previous studies of phase decomposition in binary alloys
with anisotropic elasticity have shown that the domains of hard phase are
elastically isotropic, whereas regions of soft phase are anisotropically
deformed to form a percolating network while elastic deformations are localized in
the softer regions (see 
\cite{OnukiJPhysSocJpn60,OnukiBook2002,OnukiPhysRevLet86}).

The free energy density $f(\psi,c_{\text v})$ can be expanded by $\psi$ [not $c_{\text v}$ because of the last term in equation~(\ref{f_dens1})]  giving 
\begin{equation}
\begin{split}
\frac{f(\psi,c_{\text v})}{T}&\approx \frac{E_{\text f}}{T} c_{\text v}+(1-c_{\text v})\ln (1-c_{\text v})+c_{\text v}\ln c_{\text v}+\frac{T-T_0}{T}\frac{\psi^2}{2}+\frac{\psi^4}{12}+\frac{1}{2}c_{\text v}\psi^2,
\end{split}
\end{equation}
where $T_0=Z\omega_0$ is the critical temperature in the absence of an elastic field.

The mechanical equilibrium condition 
$
\delta \mathcal{F}_0/\delta {u}_i=-\nabla_j \sigma_{ij}=0
$
 allows one to effectively exclude the elastic field by using a receipt proposed in \cite{OnukiBook2002,OnukiPhysRevB43,OnukiPhysRevLet86}, here, $\stackrel{\leftrightarrow}{\sigma}=\{ \sigma_{ij} \}$ is the elastic
stress tensor.
 To this end,  we consider 
$\mu_{21}$ and $\mu_{31}$  as small expansion parameters, namely
$|\mu_{k1}\psi|\ll L_0$ $(k=2,3)$, where $L_0=K+\mu_{20}$ is the longitudinal
elastic modulus. 
By taking into account 
$\mu_{20}=\mu_{30}$, we find:
\begin{equation}\label{expr_el}
\begin{split}
 f^{(0)}_{\text{el}}(\mathbf{u},\psi,c_{\text v}) \to  f^{(0)}_{\text{el}}(\psi,c_{\text v})&=-\frac {\alpha^2}{2L_0} \left(\psi+\frac{\beta}{\alpha}c_{\text v}\right)^2,\\
 f^{(1)}_{\text{el}}(\mathbf{u},\psi) \to f^{(1)}_{\text{el}}(\psi) &=g_2\psi |(\nabla^2_x-\nabla^2_y)\omega|^2+g_3\psi|\nabla_x\nabla_y\omega|^2.
\end{split}
\end{equation}
Here, $g_2=\mu_{21}\alpha^2/2L_0^2$ and $g_3=2\mu_{31}\alpha^2/L_0^2$ are
coefficients of elastic inhomogeneity. The potential $\omega$ is defined by the
Laplace equation   $\nabla^2 \omega=\psi-\langle\psi\rangle$  or
  $\omega=(\nabla^{2})^{-1}(\psi-\langle\psi\rangle)$,
where $(\nabla^{2})^{-1}$ is the inverse Laplacian. 

By introducing gradient energy terms $\kappa_\psi|\nabla \psi|^2$ and   $\kappa_{\text v}|\nabla c_{\text v}|^2$, and combining all the above contributions,  we obtain the total free energy of the unirradiated   system with vacancies in the form 
\begin{equation}
\mathcal{F}_0[\psi(\mathbf{r}),c_{\text v}(\mathbf{r})]=T\int_V \left[f(\psi,c_{\text v})+\kappa_\psi|\nabla \psi|^2+\kappa_{\text v}|\nabla c_{\text v}|^2\right]{\rm d}{\mathbf{r}},
\end{equation}
where, in the general case, the modified free energy density takes the form 
\begin{equation}
\begin{split}
f(\psi,c_{\text v})&=\Omega_0^{-1}[f_{\text v}(c_{\text v})+f_\psi(\psi)]+f_{c}(c_{\text v},\psi)+g_2\psi |(\nabla^2_x-\nabla^2_y)\omega|^2+g_3\psi|\nabla_x\nabla_y\omega|^2
\end{split}
\end{equation}
with  
\begin{equation}\label{f_comp}
\begin{split}
&f_{\text v}(c_{\text v})=\epsilon c_{\text v}\left(1-\lambda c_{\text v}\right)+(1-c_{\text v})\ln (1-c_{\text v})+c_{\text v}\ln c_{\text v}\,,\\ &f_\psi(\psi)=-\frac{\theta}{2}{\psi^2}+\frac{\psi^4}{12}\,,\qquad f_{c}(c_{\text v},\psi)=-\left(\gamma-\frac{1}{2}\psi\right)\psi c_{\text v}.
\end{split}
\end{equation}
Here, we introduce dimensionless energy $\lambda\equiv {\beta^2\Omega_0}/{L_0\epsilon}$ and dimensionless temperature $\theta=T_{\text C}/T-1$, where the critical temperature $T_{\text C}=T_0+\vartheta$  is defined by the  elastic contribution $\vartheta\equiv\alpha^2\Omega_0/L_0$; $\gamma\equiv-\sqrt{\epsilon\lambda\vartheta}/\Omega_0$ is responsible for  the connection between $c_{\text v}$ and $\psi$ caused by elastic contribution. By comparing the terms proportional to $\psi^2$ in the last two terms in equation~(\ref{f_comp}) it is seen that large vacancy concentration  effectively increases the temperature of the system.

In further consideration we take  into account the voids formation in an irradiated system as a process  of vacancies condensation from  vacancy-supersaturated solution. In the framework of diffuse interface model  allowing one to resolve the space and time dynamics of radiation damage,
diffusion, and microstructure and micro-chemical changes in
irradiated materials \cite{Part1,Rokkam2009},  the voids formation can be described by introducing a non-conserved phase field  $\eta$. It  plays the role of an order parameter:  $\eta=1$ within the void phase, where vacancy concentration $c_{\text v}=1$;  $\eta=0$ in the phases with $c_{\text v}=c_{\text{v}0}$. The phase field continuously varies from $\eta=0$ in the matrix phase  to $\eta=1$ in the void phase; interface between the void and matrix phase is diffuse but has a small width. 

In the phase field approach,  the shape function is defined as follows:  $h(\eta)=(\eta-1)^2(\eta+1)^2$. It varies continuously from $h(0)=1$ in the matrix phase  to  $h(1)=0$ in the void phase. Bistability of the system related to a possibility of both  matrix phase and void phase formation is described by a function $\phi(c_{\text v},\eta)=-A(c_{\text v}-c_{\text{v}0})^2\eta(\eta+2)(\eta-1)^2+B(c_{\text v}-1)^2\eta^2$,
where $A$ and $B$ are fitting parameters \cite{Rokkam2009}.

By introducing a gradient energy term $\kappa_\eta|\nabla \eta|^2$, and combining all the above contributions,  we obtain the total free energy of the irradiated system in the form 
\begin{equation}\label{totF}
\begin{split}
\mathcal{F}[\psi(\mathbf{r}),c_{\text v}(\mathbf{r}),\eta(\mathbf{r})]=T\int_V \left[
   f(\psi,c_{\text v},\eta)+\kappa_\psi|\nabla \psi|^2 
 +\kappa_{\text v}|\nabla c_{\text v}|^2+\kappa_\eta|\nabla \eta|^2\right]\rd{\mathbf{r}},
\end{split}
\end{equation}
where  the modified free energy density is of the form 
\begin{align}
f(\psi,c_{\text v},\eta)&=\Omega_0^{-1}[h(\eta) f_{\text v}(c_{\text v})+f_\psi(\psi)+\phi(c_{\text v},\eta)]+f_{c}(c_{\text v},\psi)+g_2\psi |(\nabla^2_x-\nabla^2_y)\omega|^2\nonumber\\&+g_3\psi|\nabla_x\nabla_y\omega|^2.
\end{align}

The free energy (\ref{totF}) is  used to define diffusion fluxes for both solute and vacancy  concentrations in the standard manner: $\mathbf{J}_\psi=-M_\psi\nabla\delta\mathcal{F}/\delta \psi$ and $\mathbf{J}_{\text v}=-M_{\text{v}}\nabla\delta\mathcal{F}/\delta c_{\text v}$, where $M_\psi$ and $M_{\text{v}}$ are the corresponding mobilities.
Here, we need to stress that the mobility $M_\psi$, generally,  depends on irradiation conditions and takes the form $M_\psi=M^{\text{th}}_\psi+M_\psi^{\text{irr}}$, where $M^{\text{th}}_\psi=\bar{c}_{\text A}\bar{c}_{\text B} c_{\text{v}0}D_{\text v}/T$,  $M_\psi^{\text{irr}}$ depends on  vacancies and interstitials created in the displacement cascades. As was shown in \cite{DLPS}, an expression for the irradiation-induced  mobility, $M_\psi^{\text{irr}}$, can be obtained from a stationary homogeneous vacancy and interstitial concentrations $c_{\text{v,i}}^{\text s}$ as  $M_\psi^{\text{irr}}=\frac{\bar{c}_{\text A}\bar{c}_{\text B}}{T}(c_{\text v}^{\text s}D_{\text v}+c_{\text i}^{\text s}D_{\text i})$ by considering the so-called recombination regime. It gives 
\[M_\psi^{\text{irr}}\approx\frac{\bar{c}_{\text A}\bar{c}_{\text B}}{T}\sqrt{\frac{\mathcal{K}_0\Omega_0D_{\text v}}{4\piup r_0}}\left(\sqrt{1-\varepsilon_{\text i}}+\sqrt{1-\varepsilon_{\text v}}\right).\]  
For the mobility of vacancies, we assume the standard form $M_{\text{v}}=c_{\text v}D_{\text v}/T$.
The kinetic equation corresponding to the evolution of the phase field  is
the  Allen-Cahn equation \cite{Cahn61,Cahn63} described by  the thermodynamic force $\delta\mathcal{F}/\delta\eta$ with a mobility $M_\eta=\text{const}$.
Therefore, the total set of dynamical equations takes the form 
\begin{equation}\label{totaleqs}
\begin{split}
&\partial_t\psi=\nabla\cdot M_\psi'\nabla \frac{\delta\mathcal{F}}{\delta \psi};\\
&\partial_t c_{\text v}=\mathcal{K}(1-\varepsilon_{\text v})-\mu(\rho_{\text{i,v}})(c_{\text v}-c_{\text{v}0}) -\frac{\mathcal{K}(1-\varepsilon_{\text i})(c_{\text v}-c_{\text{v}0})(1-c_{\text v})} {\delta\mu(\rho_{\text{i,v}})+(c_{\text v}-c_{\text{v}0})}+\nabla\cdot   M_{\text{v}}'\nabla \frac{\delta\mathcal{F}}{\delta c_{\text v}};\\
&\partial_t\eta=-M_\eta'\frac{\delta\mathcal{F}}{\delta \eta};\\
&  \tau_{\rho_{\text i}}\partial_{t}\frac{\rho_{\text i}}{{\rho_{\text N}}}= \mathcal{K}\varepsilon_{\text i}-\frac{\rho_{\text i}}{{\rho_{\text N}}}\left[\frac{\mathcal{K}(1-\varepsilon_{\text i})(1-c_{\text v})}{\mu(\rho_{\text{i,v}})+\delta^{-1}(c_{\text v}-c_{\text{v}0})}+(c_{\text v}-c_{\text{vL}})\right];\\
& \tau_{\rho_{\text v}}\partial_{t}\frac{\rho_{\text v}}{\rho_{\text N}}=\mathcal{K}\varepsilon_{\text v}-\frac{\rho_{\text v}}{\rho_{\text N}}\left[\frac{\mathcal{K}(1-\varepsilon_{\text i})(1-c_{\text v})}{\mu(\rho_{\text{i,v}})+\delta^{-1}(c_{\text v}-c_{\text{v}0})}-(c_{\text v}-c_{\text{v}0})\right].
\end{split}
\end{equation}

Here, we use dimensionless kinetic coefficients $M_{\psi,\text{v},\eta}'=M_{\psi,\text{v},\eta}/D_{\text v}\rho_{\text N}$ and measure space
$\mathbf{r}$ in units  $\ell=\sqrt{\kappa_\psi/T}$, $\ell\simeq (5{-}7)a$ (depending on the temperature).
In our computations we consider the case  $\varepsilon_{\text i}/\varepsilon_{\text v}>1$ with  
 $\varepsilon_{\text i}=0.9$, $\varepsilon_{\text v}=0.01$  (the relation between $\varepsilon_{\text i}$ and $\varepsilon_{\text v}$ was discussed in \cite{SemWoo}).  For other parameters, we take typical values for most of metals  with bcc structure and alloys (stainless steels):   $a=0.36$~nm, $|\mathbf{b}|=0.26$~nm, $Z=8$,  $T_{\text C}=1250$~K,  $w_0=0.0135$~eV, $\Omega_0=1.2\times 10^{-29}$~m$^{3}$,  $E_{\text f}=1.6$~eV, $E_m=1.2$~eV, $T\in[0.5T_{\text C}, T_{\text C}]$, 
 $c_{\text{v}0}\simeq \exp(-E_{\text f}/T)$, $D_{\text v}=1.29\times 10^{-6}\exp(-E_m/T)$~m$^2$/s, $\mu_2=58$~GPa, $K=168$~GPa, $\nu\simeq0.3$, $\gamma_{\text{SF}}= 0.015$~J$\cdot$m$^{-2}$, $\rho_{\text N}=10^{10}$~m$^{-2}$. We take the dose rate in the interval $\mathcal{K}_0\in[10^{-5},\,5\times10^{-2}]$~dpa/s. Its value essentially depends on the energy of incoming particle (electrons, neutrons, protons, ions), usually ranging in the intervals 300~keV--2~MeV, particle flux ($10^{12}{-}10^{15}$~cm$^{-2}$s$^{-1}$),  and  defects cross-section (from $0.18{-}5$~barn) related to the size of the incoming particle, charge of the ion (for example, Fe$^+$, Fe$^{++}$, C$^{+}$, C$^{++}$, etc.) and material properties of the target \cite{Was}. Usually, an irradiation at reactor conditions (by neutrons or electrons at energies up to $2$~MeV) corresponds to $\mathcal{K}_0\simeq 10^{-6}$~dpa/s at $T\in[600,800]$~K, whereas at irradiation in linear accelerators,  one has  $\mathcal{K}_0\simeq 10^{-4}{-}10^{-1}$~dpa/s at $T\in[800,0.9T_{\text C}]$~K at energy of ions from $300$~keV. 
To determine the dose in dpa, we use the relation $\mathcal{K}_0t=\langle\mathcal{K}\rangle t'$, where $t$ is the physical time, $t'$ is dimensionless computation time rescaled by  $D_{\text v}\rho_{\text N}$. 
 Next, we drop all primes for convenience and in the computation procedure we use: $M_\eta=1$,  $\mu_{30}/\mu_{20}=1$, 
$\mu_{21}=\mu_{31}=0.9\mu_{20}$,  $\kappa_\psi=\kappa_{\text v}=\kappa_\eta/2=1.0$, $g_2=g_3/4\cong 0.07\mu_{20}$, $\theta\in[0,1]$, $\lambda=0.1$, $\gamma=-0.1$, $A=B=\epsilon$, $\tau_{\rho_{\text v}}=10^{-7}$, $\tau_{\rho_{\text i}}=10^{-8}$, $\delta\simeq 10^{-12}$. 

We will study the void growth dynamics under conditions of vacancy generation in a crystalline matrix occurring stochastically by irradiation influence. We emulate this effect by considering the stochastic generation rate $\mathcal{K}\to\mathcal{K}(\mathbf{r},t)=\mathcal{K}\times\zeta(\mathbf{r},t)$, where $\mathcal{K}$ is a constant fixed by irradiation conditions and associated with a single cascade event, whereas $\zeta$ is a dimensionless random number ranging in the interval  [0,1] with the uniform distribution \cite{Rokkam2009}. It is used to determine the source strength assigned at the random point $\mathbf{r}$ in the matrix phase in each time instant which is fixed by considering the value of the order parameter for the void phase $\eta$ \cite{Rokkam2009}.  

\section{Linear stability analysis} \label{3}

The first step in studying the system dynamics lies in the stability analysis of steady states. We define stationary values of vacancy concentration and phase field ($c_{\text v}^{\text s},\eta^{\text s}$), by considering a homogeneous system at $\partial_t{c_{\text v}}=\partial_t \eta=0$ by taking into account $c_{\text v}\in[c_{\text{v}0}, 1]$ and $\eta\in[0, 1]$. As far as the solute concentration obeys the conserved dynamics, for its stationary value we take its initial value $\psi_0$. Sink densities are slow modes due to small values of both scaling coefficients $\tau_{\rho_{\text i}}$ and $\tau_{\rho_{\text v}}$. It allows us to consider $\rho_{\text i}/\rho_{\text N}$ and $\rho_{\text v}/\rho_{\text N}$ as constants and control the effect of sinks by taking fixed values for the total sinks strength, $\mu$.
Next, acting in the standard manner, we consider the stability of small perturbations 
$\delta c_{\text v}(\mathbf{r},t)= c_{\text v}(\mathbf{r},t)-c_{\text v}^{\text s}$, $\delta \eta(\mathbf{r},t)= \eta(\mathbf{r},t)-\eta^{\text s}$, $\delta \psi(\mathbf{r},t)=\psi-\psi_0$, where $\delta c_{\text v}(k,t), \delta \eta(k,t),\delta\psi(k,t)\propto \re^{\Lambda(k)t}$, here, $\Lambda$ is the stability exponent, $k=|\mathbf{k}|$ is the wave number.

\begin{figure}[!t]
\centering
a)\raisebox{-0.8\height}{\includegraphics[width=0.47\textwidth]{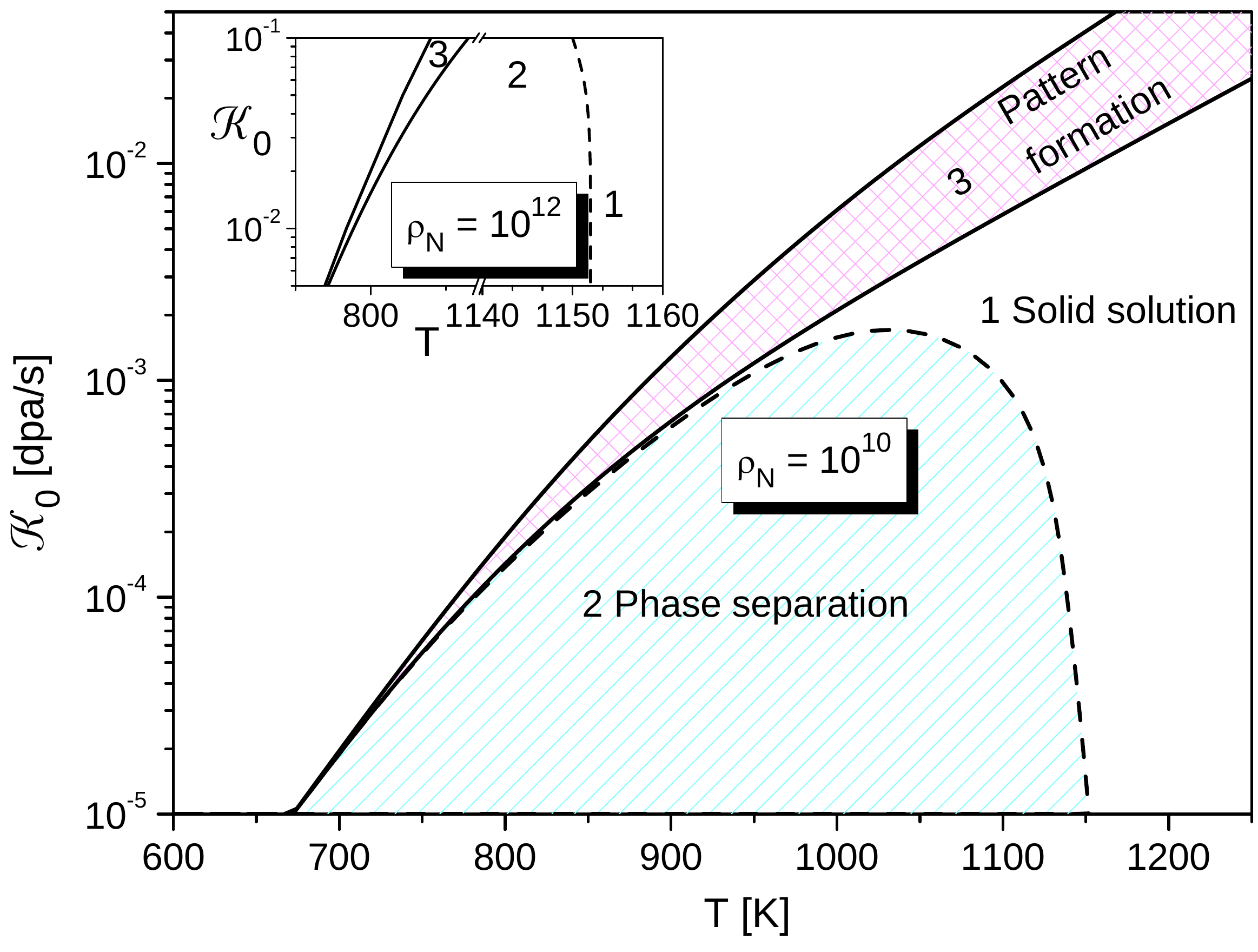}}
b)\raisebox{-0.8\height}{\includegraphics[width=0.47\textwidth]{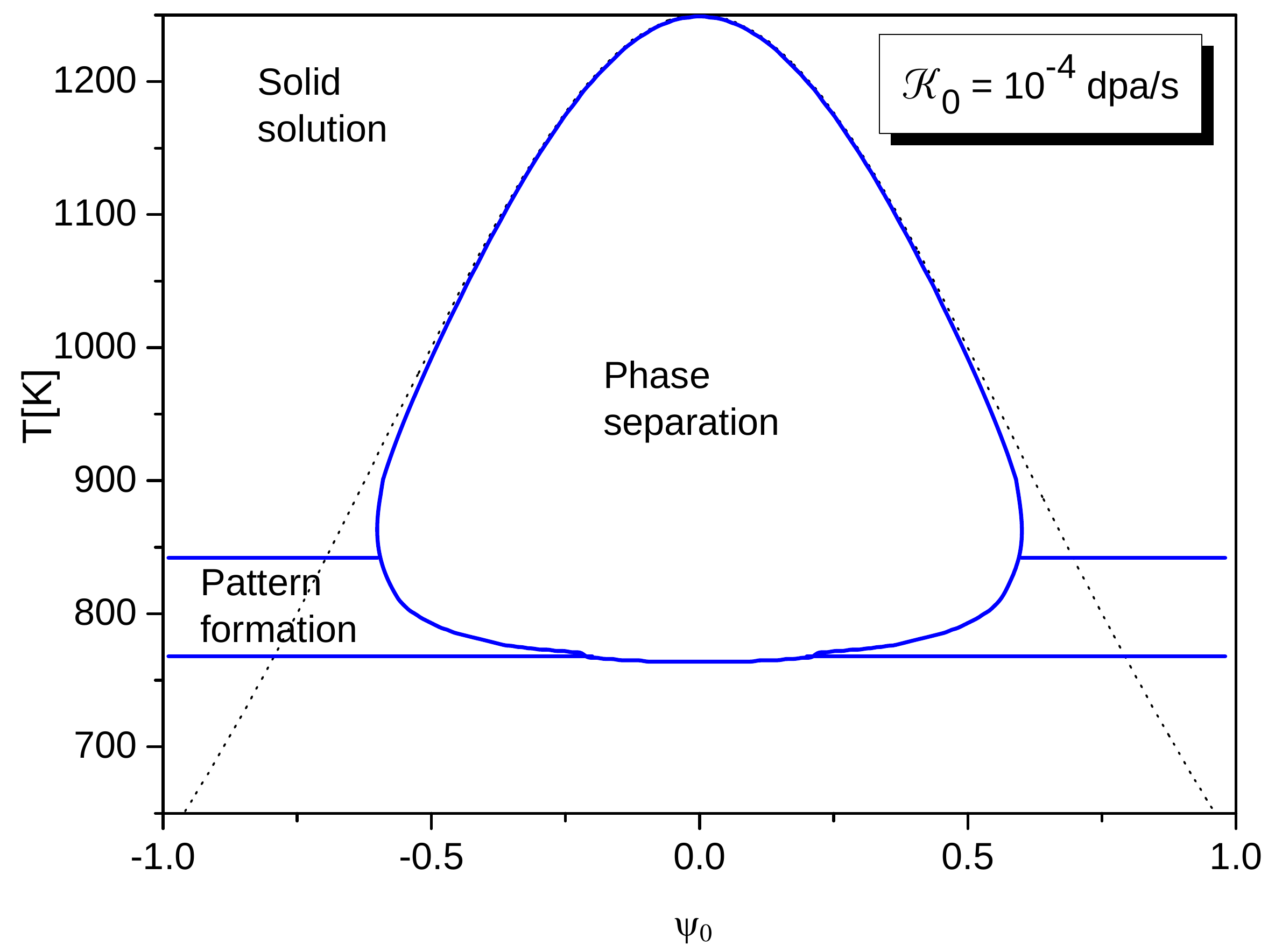}}\\ 
c)\raisebox{-0.8\height}{\includegraphics[width=0.47\textwidth]{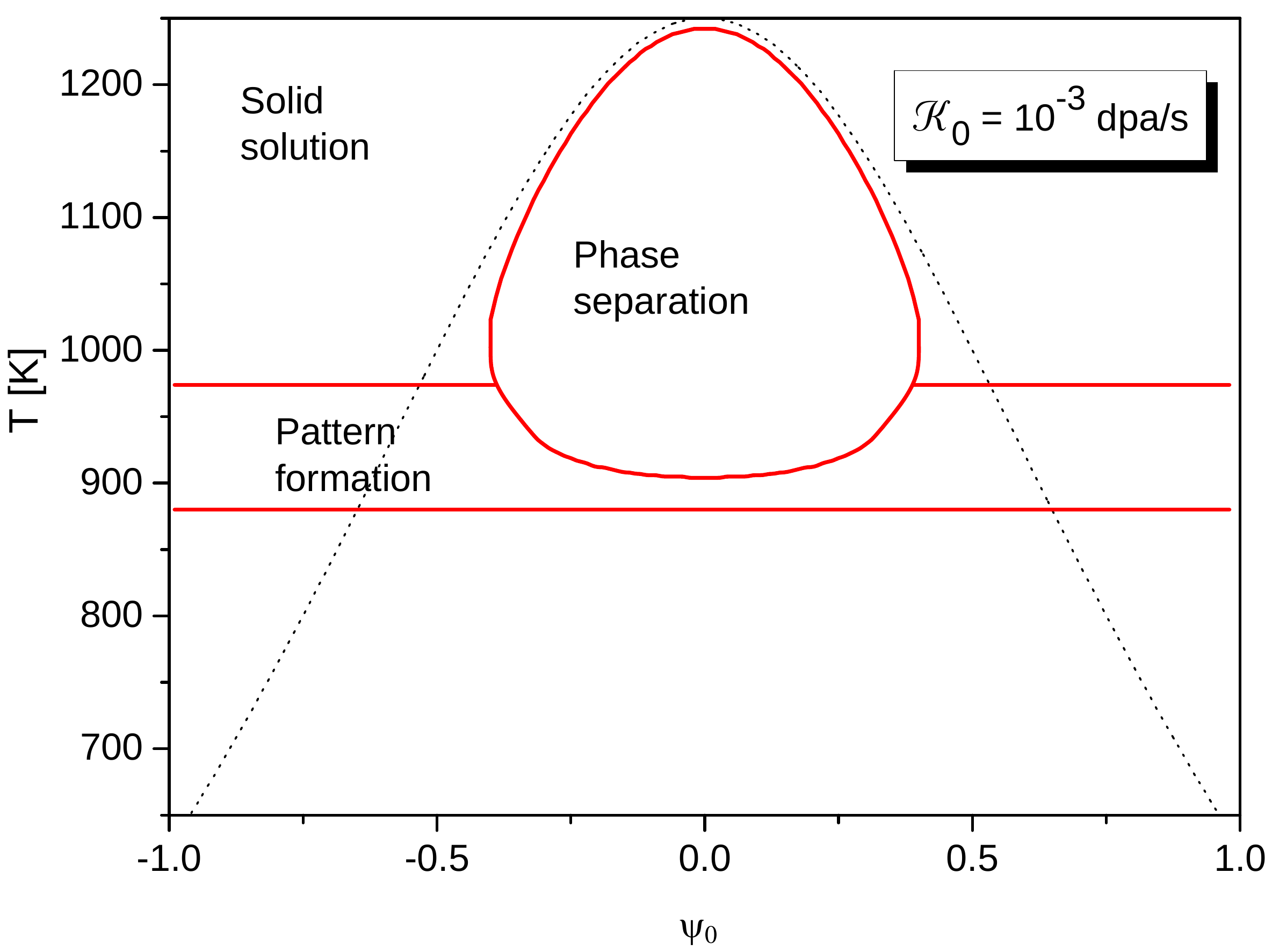}}
\caption{(Colour online) (a) Phase diagram in the plane $(T,\mathcal{K}_0)$ obtained at $\psi_0=0.3$, $\rho_{\text N}=10^{10}$~m$^{-2}$, insertion relates to $\rho_{\text N}=10^{12}$~m$^{-2}$. Plots (b) and (c) are phase diagrams in the plane $(\psi, T)$ obtained at  $\rho_{\text N}=10^{10}$~m$^{-2}$ at different dose rates; dot line relates to unirradiated system.
\label{fig_phd}}
\end{figure}

Studying the behaviour of the maximal  stability exponent we get  phase diagrams shown in  the plane $(T, \mathcal{K}_0)$ in figure~\ref{fig_phd}~(a) at $\psi_0=0.3$ and in figure~\ref{fig_phd}~(b),~(c)  in  the plane $(\psi,T)$ at different dose rates. 
Let us consider the diagram in figure~\ref{fig_phd}~(a) in detail. 
It is seen that the whole plane  is divided into three domains corresponding to phase separation, patterning, and disordered configuration related to solid solution. 
Here, the upper solid line defines the minimal temperature $T^*$ of the irradiated alloys depending  on the damage rate $\mathcal{K}_0$. The dependence $T^*(\mathcal{K}_0)$ is obtained at $c_{\text v}^{\text s},\eta^{\text s}\to 1$, $\psi=\psi_0$. It bounds the physically realized values of the temperature of irradiated system: physically possible values of $T$ and $\mathcal{K}_0$ are below the line $T^*(\mathcal{K}_0)$. The domain of solid solution is characterized by three negative  eigenvalues $\Lambda(k)$. Here, all spatial disturbances relax due to the emergence of nonequilibrium vacancies at  the structural disorder formation with non-supersaturated vacancy ensemble. 
In the domain related to phase separation,  the maximal $\Lambda(k)$ behaves in a typical manner  for phase decomposition: $\Lambda(k=0)=0$, $\Lambda(0<k<k_{\text c})>0$, where $k_{\text c}$ is the critical wave number. Here, spatial instability is realized from long length modes related to $k=0$; a threshold for unstable modes is limited by  $k=k_{\text c}$; the most unstable mode characterized by $k=k_{\text m}$ is defined by position of the maximum of $\Lambda(k)$. 
In the  domain of pattern formation, the maximal stability exponent $\Lambda(k)$ takes up positive values in the interval $k_{\text{c1}}<k<k_{\text{c2}}$, where $k_{\text{c1, c2}}\ne 0$: here $\Lambda(k=0)=0$, $\Lambda(0<k<k_{\text{c1}})<0$,  $\Lambda(k>k_{\text{c2}})<0$. The corresponding most unstable mode relates to the period of realized patterns.  From the obtained dependencies it follows that a growth in the dose rate producing large amount of vacancies leads to supersaturation of these point defects which are capable of organizing into patterns of small wavelengths. Such vacancy structures represent small vacancy clusters, defect walls,  voids, as usual. An insertion in figure~\ref{fig_phd} illustrates a change in the stability diagram with an increase in the number of sinks ($\rho_{\text N}=10^{12}$~m$^{-2}$). It is seen that the domain of temperature values where  phase decomposition takes place broadens, and the patterning is observed at a narrow interval of elevated dose rates. At a large dislocation density, the number of sinks absorbing nonequilibrium  vacancies increases and, as a result, vacancy concentration becomes small that suppresses the disordering processes.

By fixing the dose rate one can obtain a phase diagram shown in  the plane $(\psi,T)$ in figures~\ref{fig_phd}~(b),~(c) at different dose rates. Here, the dot line relates to unirradiated system: above the dot line, one has solid solution, below this line, phase decomposition takes place. If the system is subjected to irradiation, then the domain of phase decomposition becomes bounded at low temperatures. This irradiation-induced effect is caused by a large amount of vacancies produced at irradiation leading to a structural disorder with an increased atomic mobility. By comparing   figure~\ref{fig_phd}~(b) and figure~\ref{fig_phd}~(c) it follows that at an elevated dose rate the size of the domain of phase decomposition decreases from both low and high temperatures. 
Similar reentrance and the corresponding irradiation-induced effects  were discussed previously in \cite{Martin,Vaks,MTMA96} where it was shown that an emergence of the disordered phase at low temperatures is a result of ballistic mixing of atoms caused by structural disorder formation.   By considering a shift of the domain of pattern formation at the dose rate increase, one finds that patterning  is observed at higher temperatures at larger dose rates. It follows that alloys with a small concentration of alloying elements will not manifest a phase decomposition at irradiation, and only patterning caused by the formation of defect clusters, walls and voids becomes possible. It is important to stress that since we consider the  model where three main modes $\psi$, $c_{\text v}$ and $\eta$ are effectively coupled, the pattern formation relates to the formation of defect structures leading to a rearrangement of atoms of the solute with the formation of patterned structure of the alloy. Bottom solid line defines the minimal temperature of the irradiated alloy, $T^*$, as was discussed above. 

The calculated phase diagrams relate well to the diagrams obtained in \cite{DLPS,Nastar}  and to the experimental observations \cite{phd_WA97} of patterning, phase decomposition and disordered microstructure of binary alloy systems irradiated by heavy ions. There is a  qualitative agreement with the data shown in \cite{GWZ} for  pure Ni and Cu irradiated by self-ions, neutrons and protons.    Moreover, the obtained results qualitatively correspond to the results of the work  \cite{EB2000} devoted to a numerical study of the patterning in binary alloys subjected to irradiation. It should be noted that in the cited numerical studies, an assumption of ballistic mixing was used for a solute concentration in order to model the structural disorder produced at irradiation. In our case,  no ballistic mixing was incorporated into the model. Here, the structural disorder emerges as a result of stochastic defects production in cascades leading to a mixing of atoms of the solute due to the coupling of vacancy concentration with solute concentration.

\newpage

\section{Modelling} \label{4}

In this section, we consider the dynamics of the system  numerically on a two-dimensional square lattice with $N\times N$ grid points  with $N=256$ and the mesh size $\Delta x=\Delta y=0.5$; the linear size of the system is $L_{\text s}=N\ell$, $\ell\approx 2$~nm, $L_{\text s}\approx 0.5$~{\textmu}m. Time step in our simulations is $\Delta t=10^{-3}$.   
Boundary conditions are  periodic
and the initial configuration is given by $\langle\psi(\mathbf{r},0)\rangle=\psi_0$, $\langle c_{\text v}(\mathbf{r},0)\rangle=10^{-3}$, $\langle \eta(\mathbf{r},0)\rangle=0.0$ with dispersions: $\langle (\delta \psi(\mathbf{r},0))^2\rangle=0.01\psi_0$, $\langle (\delta c_{\text v}(\mathbf{r},0))^2\rangle=0.01c_{\text{v}0}$, $\langle (\delta \eta(\mathbf{r},0))^2\rangle=0.0$; $\rho_{\text v}(0)=\rho_{\text i}(0)=0$. 

\begin{figure}[!b]
\centering
\includegraphics[width=0.5\textwidth]{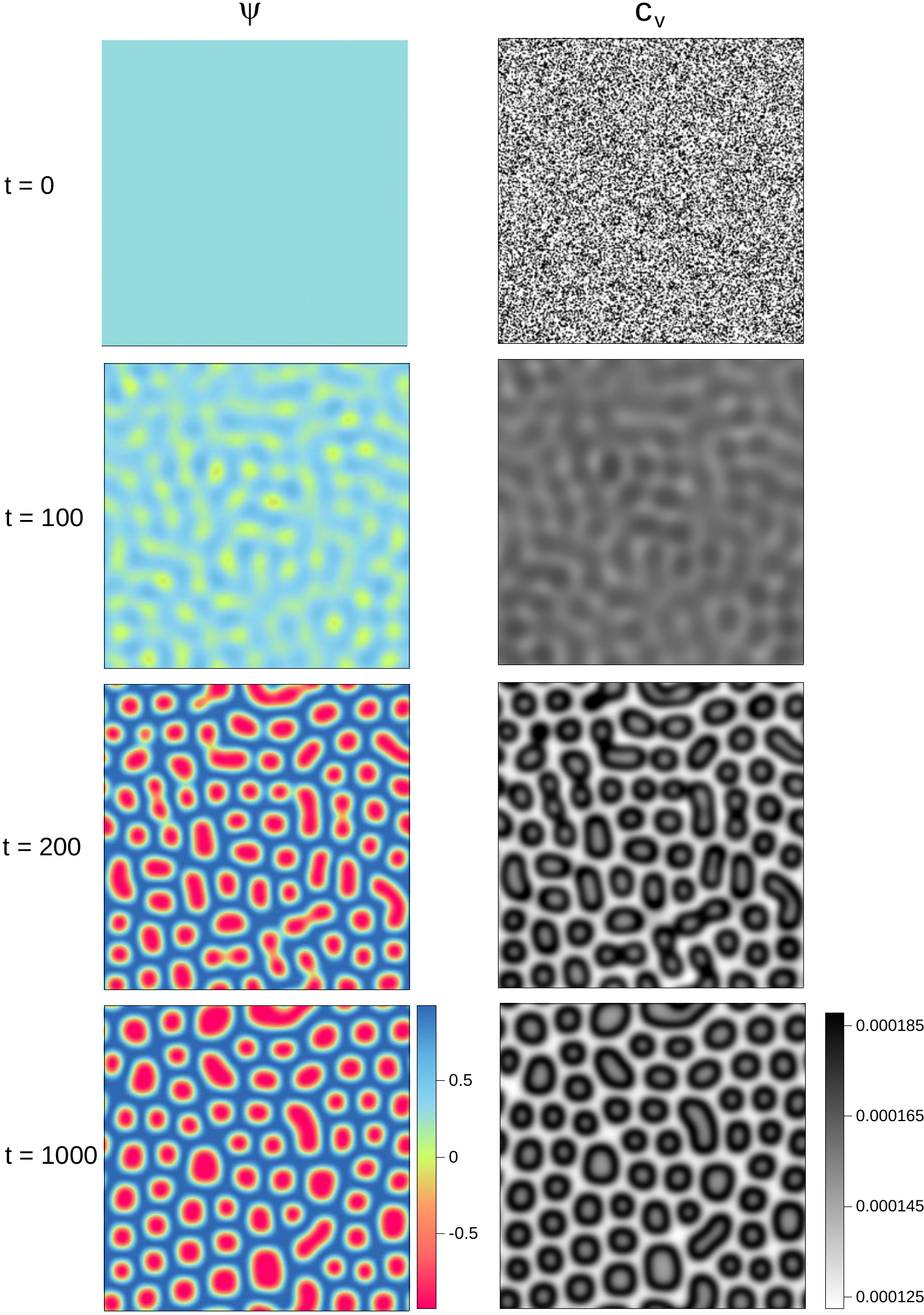}
\caption{(Colour online) Snapshots of the unirradiated system evolution. \label{fig_snapshots}}
\end{figure}
\begin{figure}[!t]
\centering
\includegraphics[width=0.5\textwidth]{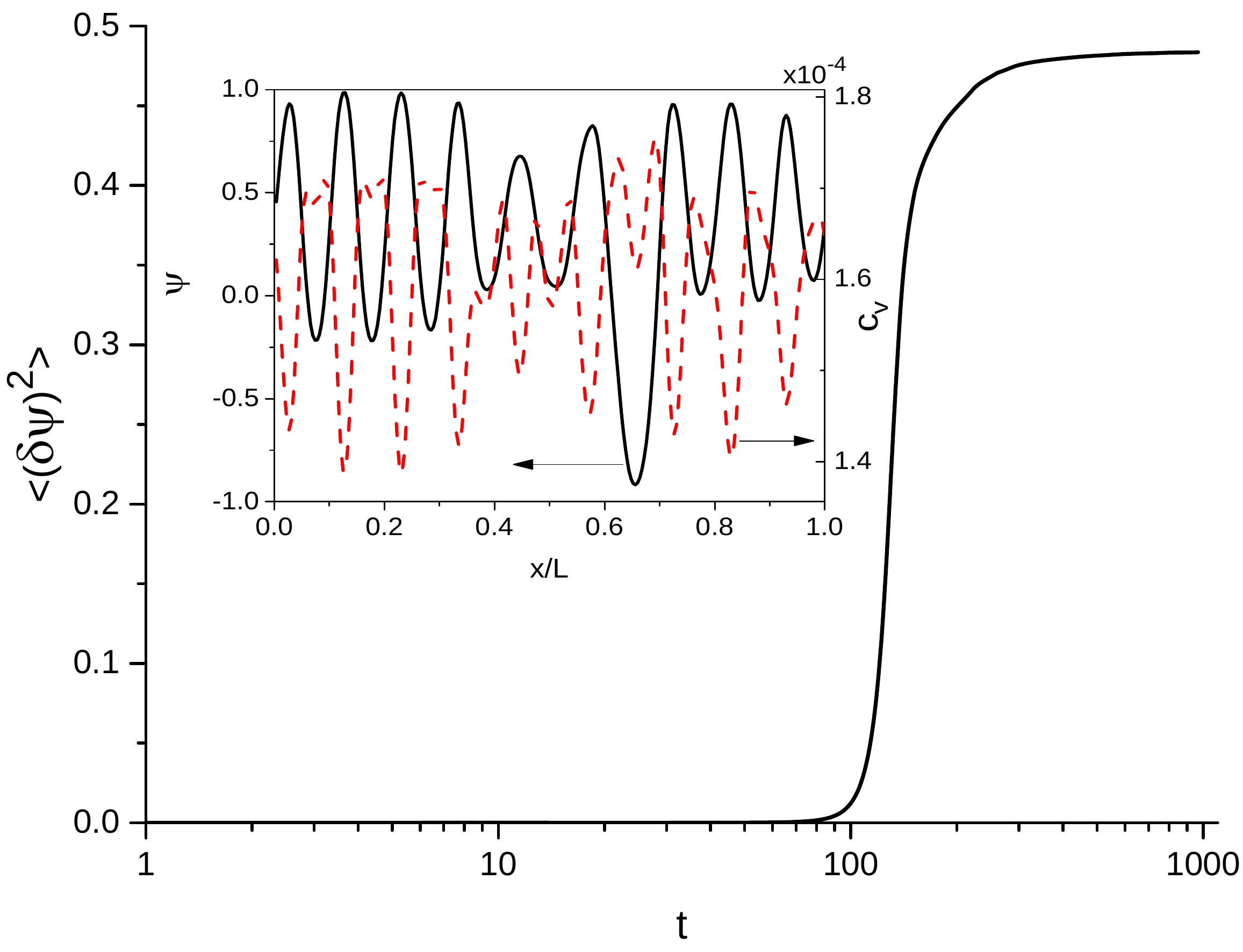}
\caption{(Colour online) Dynamics of the solute  dispersion $\langle(\delta \psi)^2\rangle$ at phase decomposition of unirradiated alloy and typical profiles for the solute concentration and vacancy concentration are shown in insertion at $t=1000$.     \label{fig_prep}}
\end{figure}

\subsection{Sample preparation}
By considering an  unirradiated system, one should take into account that no new entities (defects, atoms) can be produced, and therefore the mass conservation law $c_{\text A}+c_{\text B}+c_{\text v}=1$ holds.
To prepare a sample (target) for future irradiation, we start from a disordered solid solution  and consider the dynamics of three fields according to the following equations:
\begin{equation}\label{model}
\begin{split}
&\partial_t\psi=\nabla\cdot M_\psi\nabla \frac{\delta\mathcal{F}}{\delta \psi}+\xi_\psi(\mathbf{r},t),\\
&\partial_t c_{\text v}=\nabla\cdot M_{c_{\text v}}\nabla \frac{\delta\mathcal{F}}{\delta c_{\text v}}+\xi_{c_{\text v}}(\mathbf{r},t),\\
&\partial_t\eta=-M_\eta\frac{\delta\mathcal{F}}{\delta \eta}+\xi_\eta(\mathbf{r},t).
\end{split}
\end{equation}
In a system of equations (\ref{model}) we introduce internal fluctuation sources ($\xi_\psi$, $\xi_{c_{\text v}}$, $\xi_\eta$) which are of thermal character. They have the following Gaussian properties: $\langle \xi_\psi(\mathbf{r},t)\rangle=\langle \xi_{c_{\text v}}(\mathbf{r},t)\rangle=\langle \xi_\eta(\mathbf{r},t)\rangle=0;$
$\langle \xi_\psi(\mathbf{r},t)\xi_\psi(\mathbf{r}',t')\rangle=-\nabla^2M_\psi\delta(t-t')\delta(\mathbf{r}-\mathbf{r}')$,  $\langle \xi_{c_{\text v}}(\mathbf{r},t)\xi_{c_{\text v}}(\mathbf{r}',t')\rangle=M_{c_{\text v}}\delta(t-t')\delta(\mathbf{r}-\mathbf{r}')$,  $\langle \xi_\eta(\mathbf{r},t)\xi_\eta(\mathbf{r}',t')\rangle\\=M_{\eta}\delta(t-t')\delta(\mathbf{r}-\mathbf{r}')$.

Snapshots illustrating the dynamics of phase decomposition and the corresponding vacancies rearrangement are shown in figure~\ref{fig_snapshots} at an initial condition  $\psi_0=0.3$ and at $T=500$~K. 
The dynamics of dispersion of the solute concentration $\langle(\delta\psi)^2\rangle$ is shown in figure~\ref{fig_prep}. A growth in $\langle(\delta\psi)^2\rangle$ indicates the formation of an ordered configuration where the domains of two phases emerge during the system evolution.
The obtained results correspond well to previous studies of a vacancy rearrangement in binary systems with soft and hard phases \cite{UJP2016,REDS2016}. It was shown that at phase decomposition, the above rearrangement of vacancies is a result of Kirkendall effect occurring in the systems with different self-diffusivities of atoms of two sorts. This effect is clearly seen in figure~\ref{fig_prep}, where profiles of both solute and vacancy concentrations are plotted in the insertion. It follows that the connection of vacancy concentration  with elastic deformation leads to the location of vacancies  in the  ``soft'' phase. If there is no connection between vacancy concentration and elastic deformation (vacancies do  not change the lattice constants of the alloy), then only two states for a vacancy ensemble are possible: a state related to interfaces described by  $c_{\text v}\simeq c_{\text{v}0}+c^{\text{intf}}_{\text v}$, where  $c^{\text{intf}}$ is the vacancy concentration in the middle of the phase interface,  and $\psi\simeq0$, and a state corresponding to the domains of two phases, where   $c_{\text v}\simeq c_{\text{v}0}-c^{\text{intf}}_{\text v}$ and $\psi\ne0$. In the case of $\lambda\ne0$ (vacancies affect the lattice constant by changing the elastic field locally), there are three states for a vacancy subsystem: the first one  relates to interfaces  ($c_{\text v}\gtrsim c_{\text{v}0}+c^{\text{intf}}_{\text v}$,  $\psi\simeq0$), the second state  corresponds to the domain of a hard phase ($c_{\text v}<c_{\text{v}0}-c^{\text{intf}}_{\text v}$,  $\psi>0$), the third state relates to a soft phase with ($c_{\text v}\simeq c_{\text{v}0}$,  $\psi<0$). The main statistical properties of the target prepared in the discussed way can be found in \cite{PhysA2017}.

\subsection{Irradiation}

By considering the irradiated systems, one should take into account that mass conservation here is held only for the field $\psi$, obeying the conserved dynamics, i.e, $\int\psi(\mathbf{r},t){\rm d}\mathbf{r}=\psi_0=\text{const}$. At the same time, vacancies are constantly produced and react with the other defects. Therefore, mass conservation law relevant to the system in a thermodynamic equilibrium can be violated as far as the irradiated systems are the \emph{driven} ones, where the irradiation deposits energy and ``mass'' into a system. 

\begin{figure*}[!t]
\centering
 \includegraphics[width=0.8\textwidth]{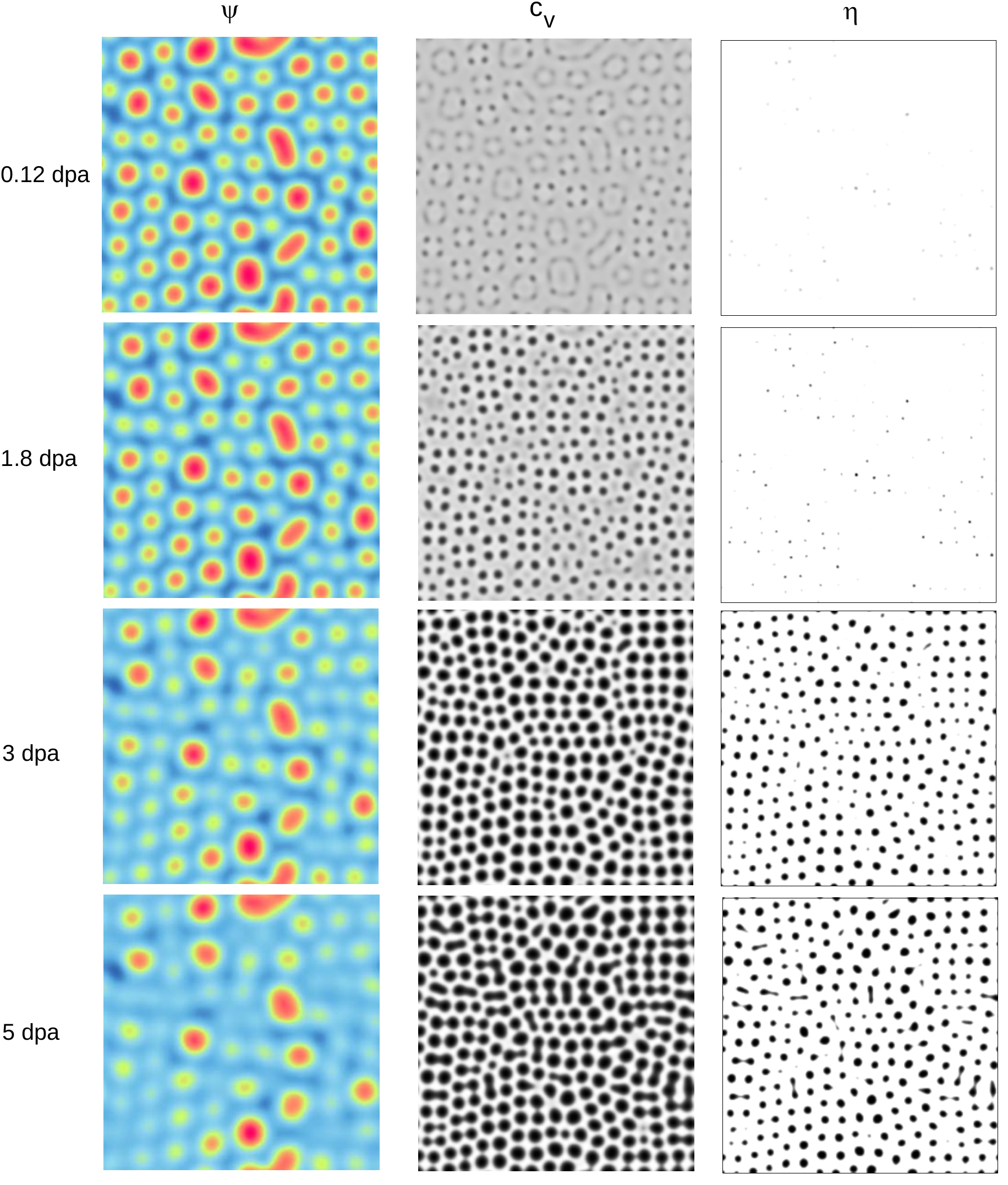}
\caption{(Colour online)  Snapshots of the system microstructure evolution during irradiation at $\mathcal{K}_0\simeq 2\times10^{-2}$~dpa/s, $T\simeq1136$~K.
\label{fig_snap_irr}}
\end{figure*}
To study the system dynamics at irradiation, we numerically solve the system (\ref{totaleqs}) by taking the configuration from figure~\ref{fig_snapshots} at $t=1000$ as an initial condition.
Snapshots of the system evolution at  $\mathcal{K}_0\simeq2\times 10^{-2}$~dpa/s, $T\simeq1136$~K are shown in figure~\ref{fig_snap_irr}.  It is seen that  the production of vacancies increasing the solute mobility leads to a disordering of the composition field $\psi$ related to dissolution of the precipitates. At the same time, these  vacancies segregate at interfaces and in the ``soft'' phase after supersaturation. This effect is caused by  interaction of vacancies through elastic fields. It leads to the formation of vacancy clusters at interfaces having a large curvature, where elastic deformation of the lattice is large. Vacancy clusters are also formed inside the ``soft'' precipitates, but their formation inside such a phase is less probable (see snapshot at $0.12$~dpa). With the dose growth, the number of vacancies in clusters increases and if the corresponding supersaturation is sufficient, then voids are formed (see snapshot at $2$~dpa). These voids grow and their number increases. At large doses (see snapshot at $5$~dpa),  the already existing voids continue to grow by absorbing vacancies from the bulk. Large voids grow at the expense of small
ones in their vicinity according to the Ostwald ripening scenario, and their number decreases \cite{Ostwald0,Ostwald1,Ostwald}.
Here, one can observe a formation of a diffuse void-matrix interface. From the formalism of the phase field approach it follows that an emergence of this interface is a result of a competition between the gradient and bulk energy terms in the free energy functional: the gradient energy terms are responsible for the curvature of the interface, whereas  its formation is described by the terms corresponding to  a free energy decrease \cite{Rokkam2009}. The void shape
development is governed by different mechanisms related to: surface
energy anisotropy, growth anisotropy of void surfaces,  preferential
adsorption of atoms on certain surfaces of voids during
irradiation \cite{Chen73,Anisvoids2016}. In our study,  we consider the simplest case of isotropic void surface energy. 
\begin{figure*}[!t]
\centering
a)\raisebox{-0.8\height}{ \includegraphics[width=0.45\textwidth]{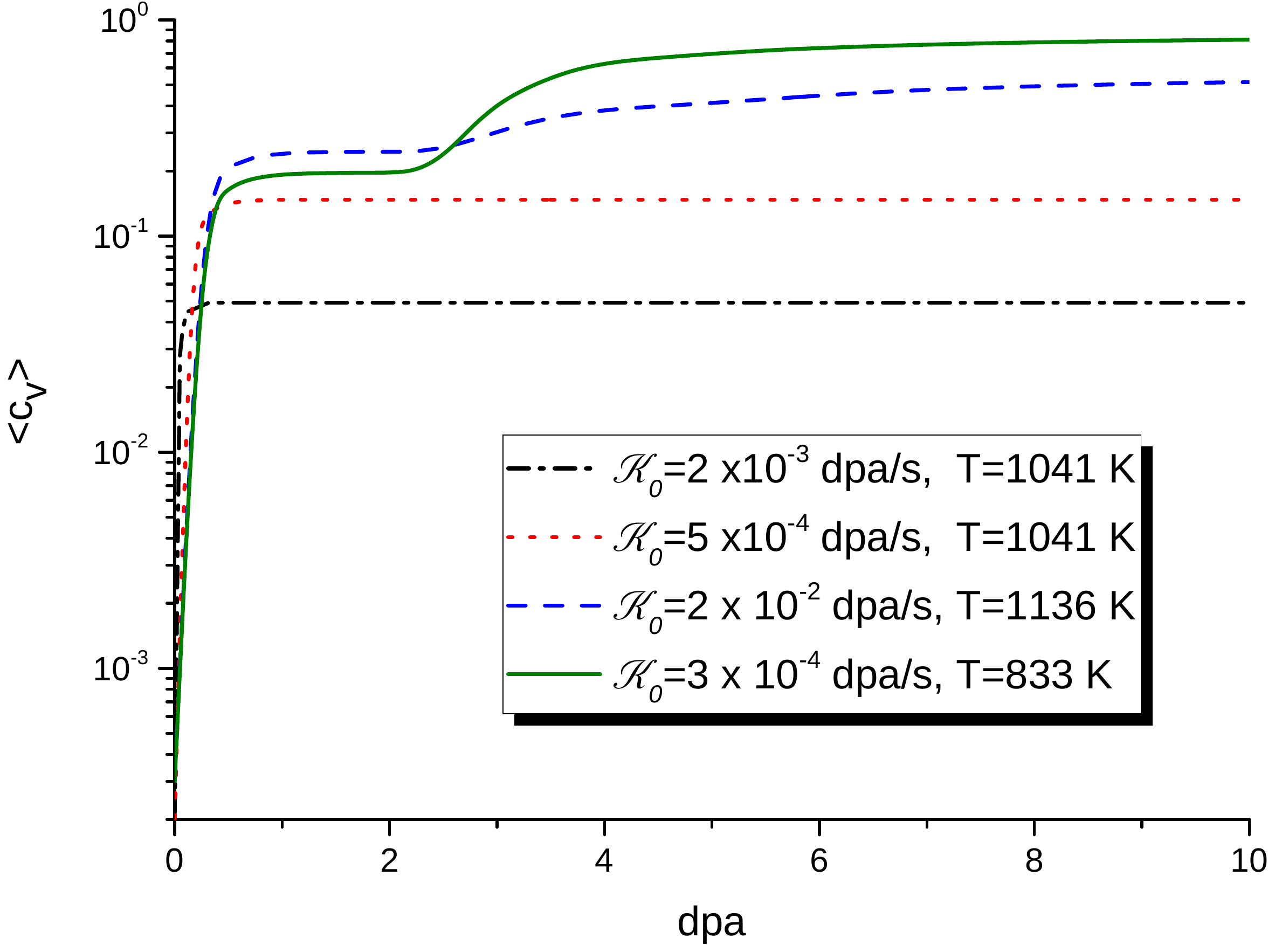}}\hfill
b)\raisebox{-0.8\height}{ \includegraphics[width=0.45\textwidth]{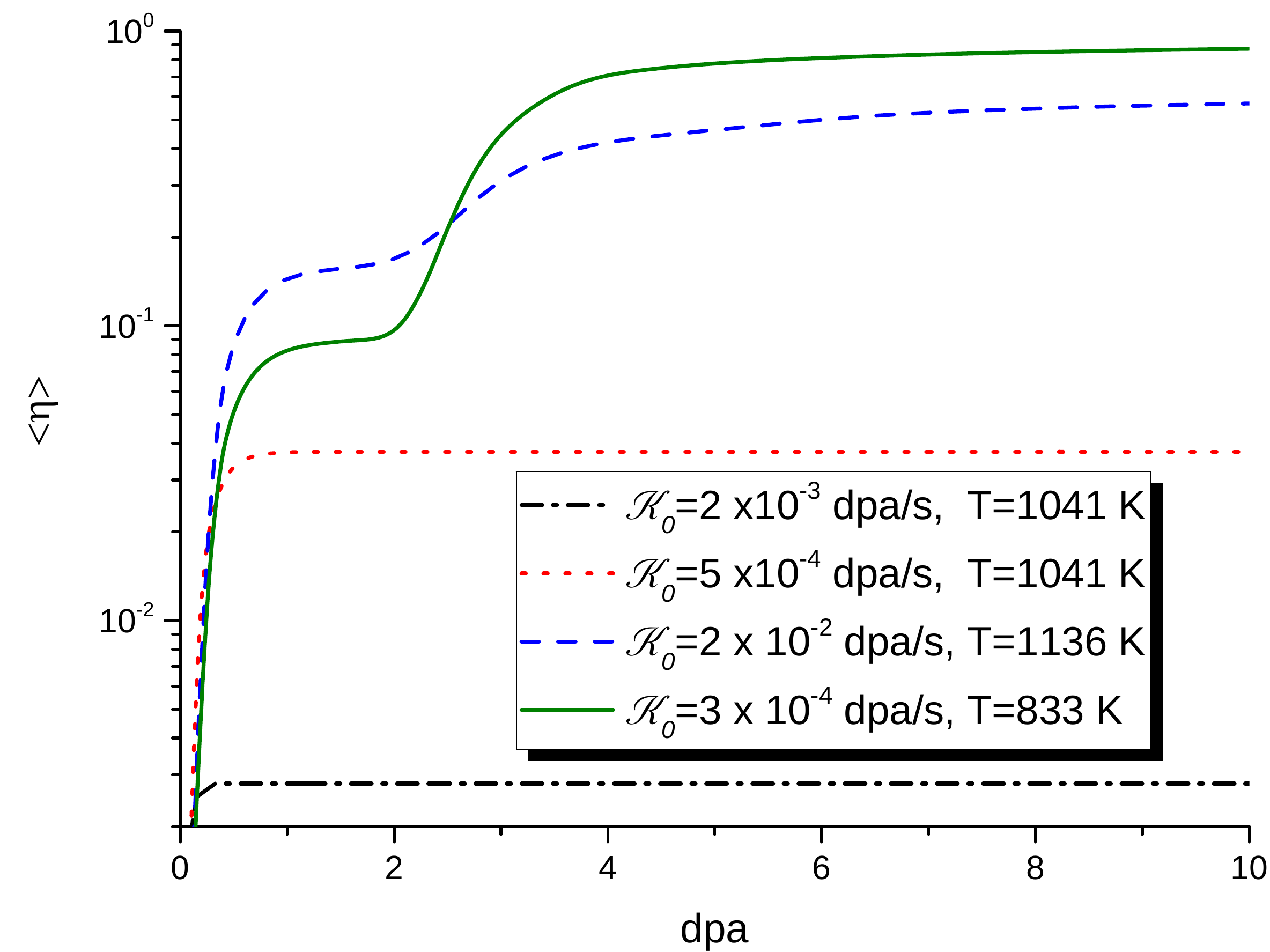}}\\
c)\raisebox{-0.8\height}{ \includegraphics[width=0.45\textwidth]{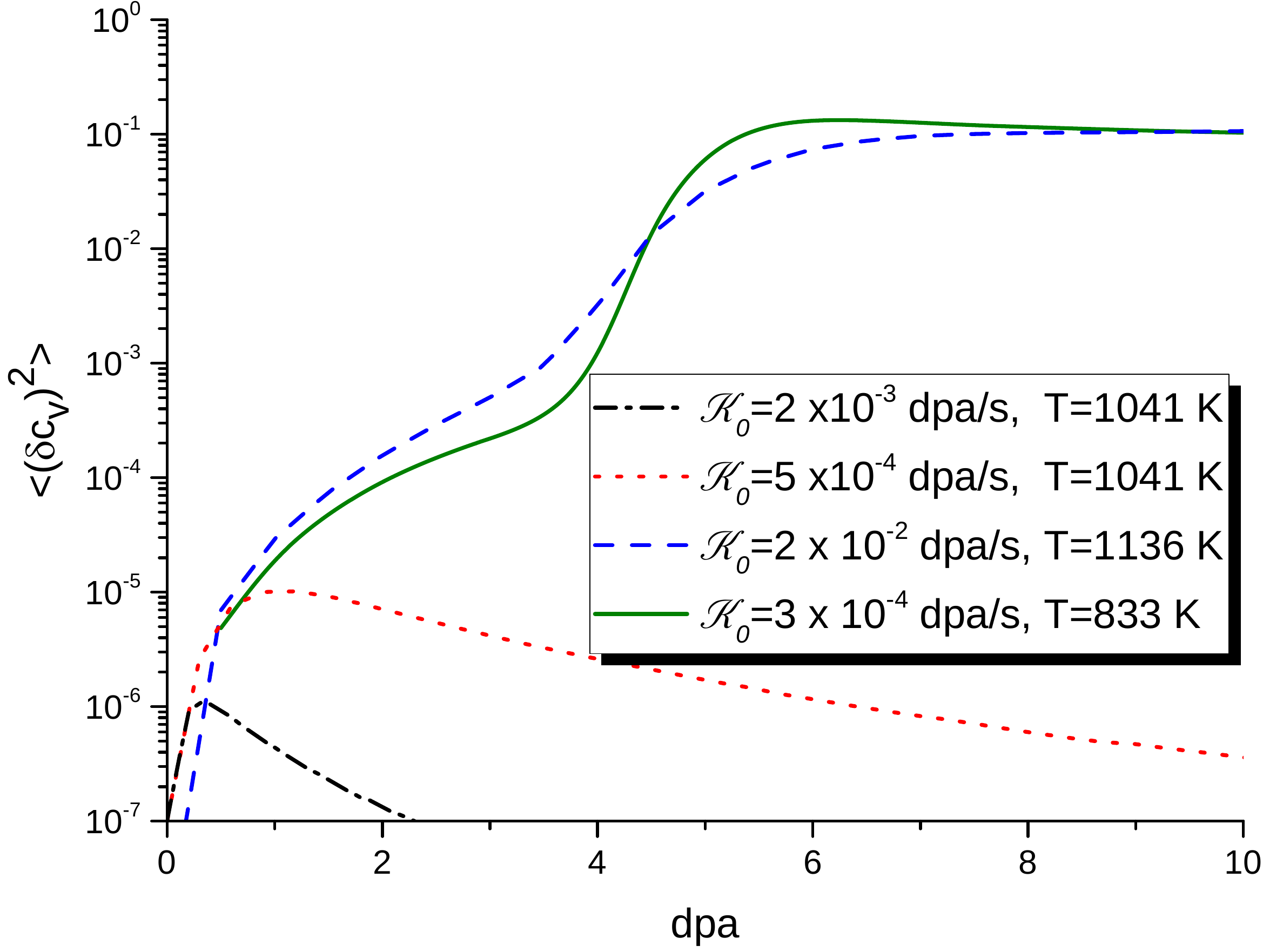}}\hfill
d)\raisebox{-0.8\height}{ \includegraphics[width=0.45\textwidth]{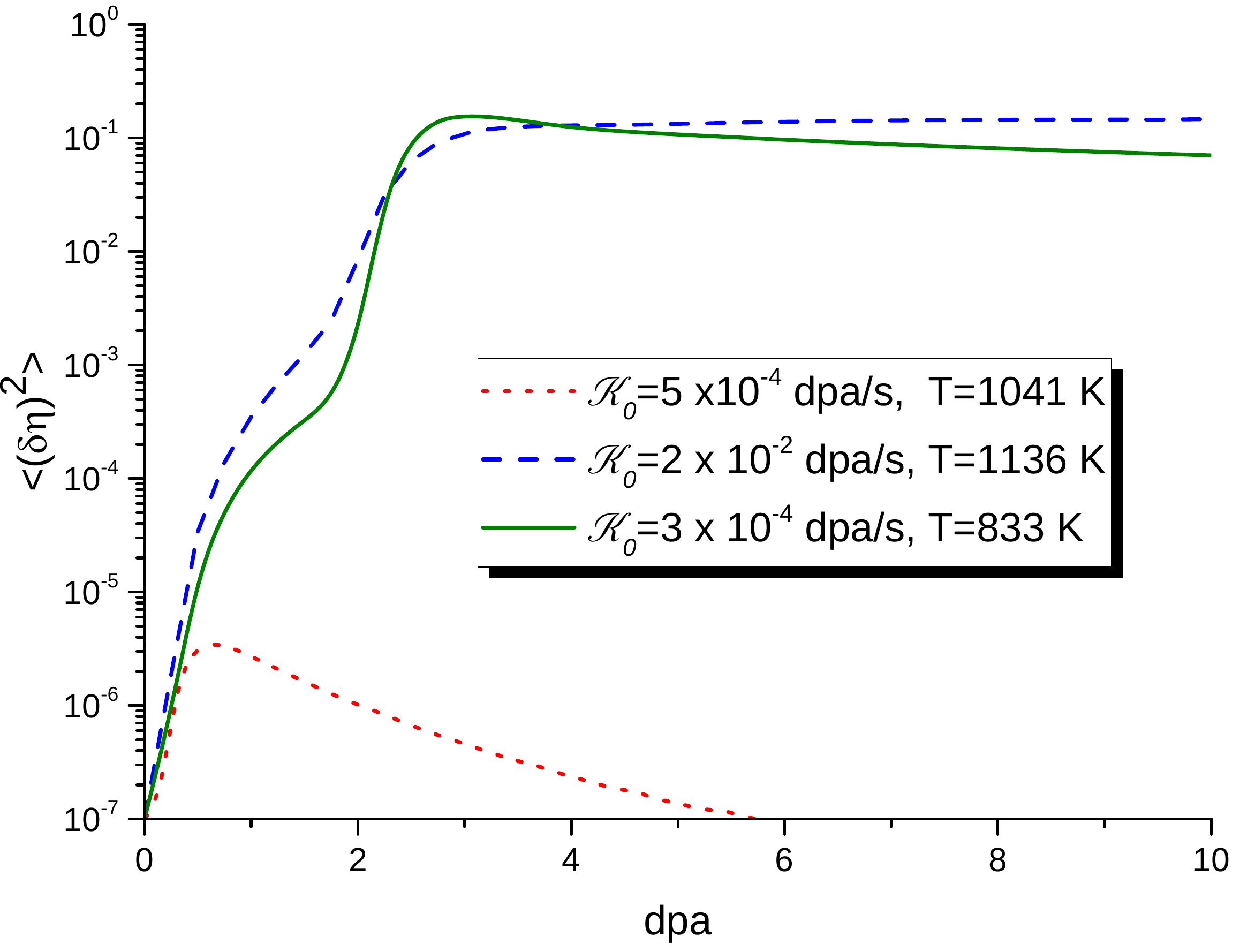}}\\
e)\raisebox{-0.8\height}{ \includegraphics[width=0.45\textwidth]{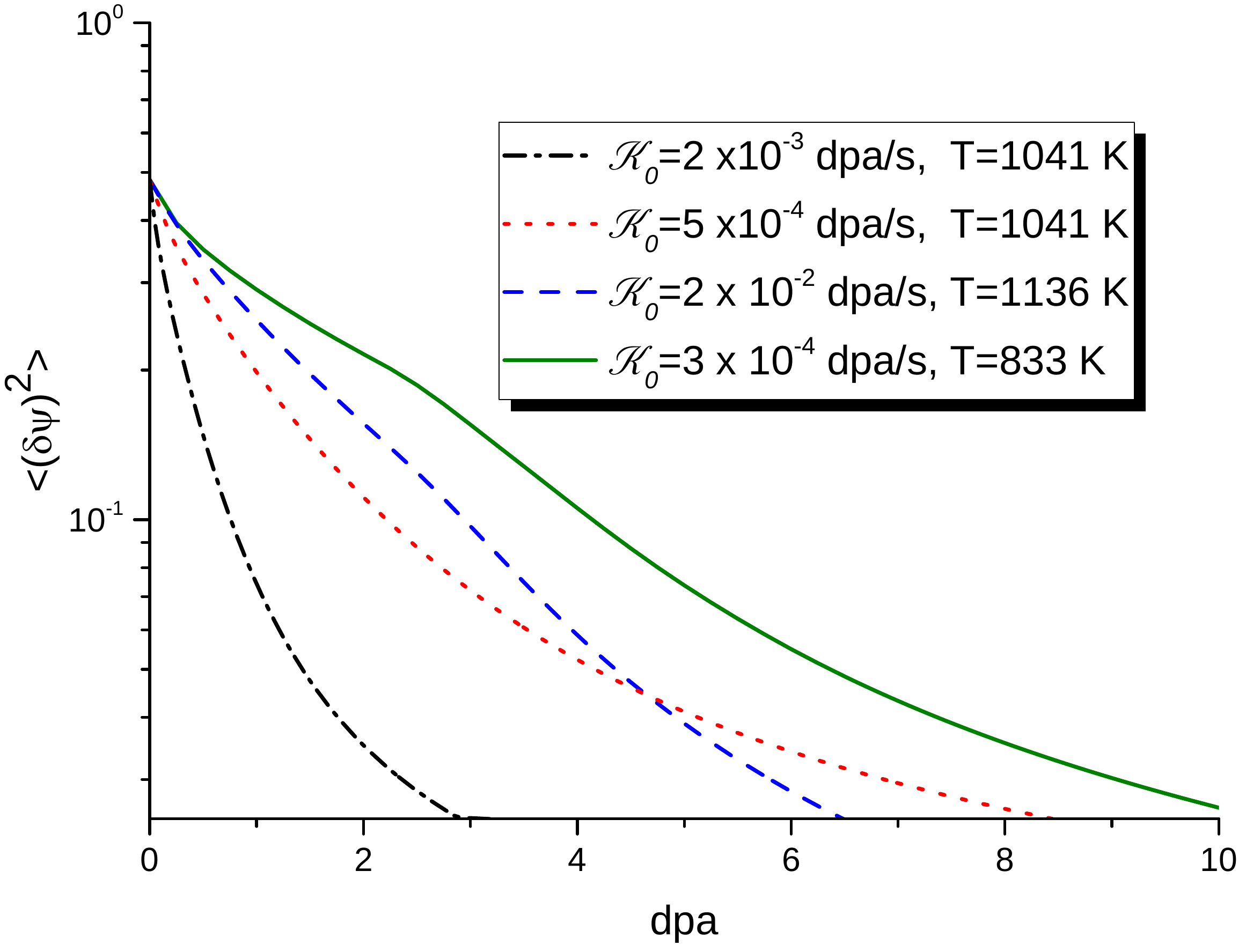}} 
\caption{(Colour online) Dynamics of averaged vacancy concentration (a),  phase field (b). Plots (c)--(e) correspond to the dynamics of  dispersions of vacancy concentration, phase field, and solute concentration, respectively.
\label{fig_irr}}
\end{figure*}

\begin{figure*}[!t]
\centering
$\psi$\hspace{35mm} $c_{\text v}$\hspace{35mm}$\eta$\\
a)\raisebox{-0.8\height}{ \includegraphics[width=0.78\textwidth]{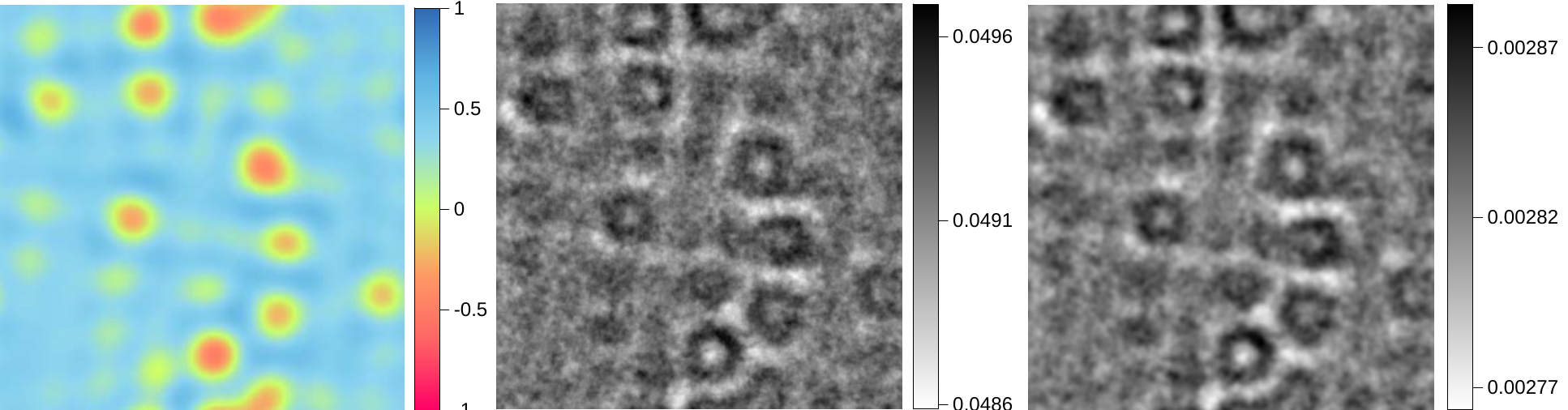}}\\
b)\raisebox{-0.8\height}{ \includegraphics[width=0.78\textwidth]{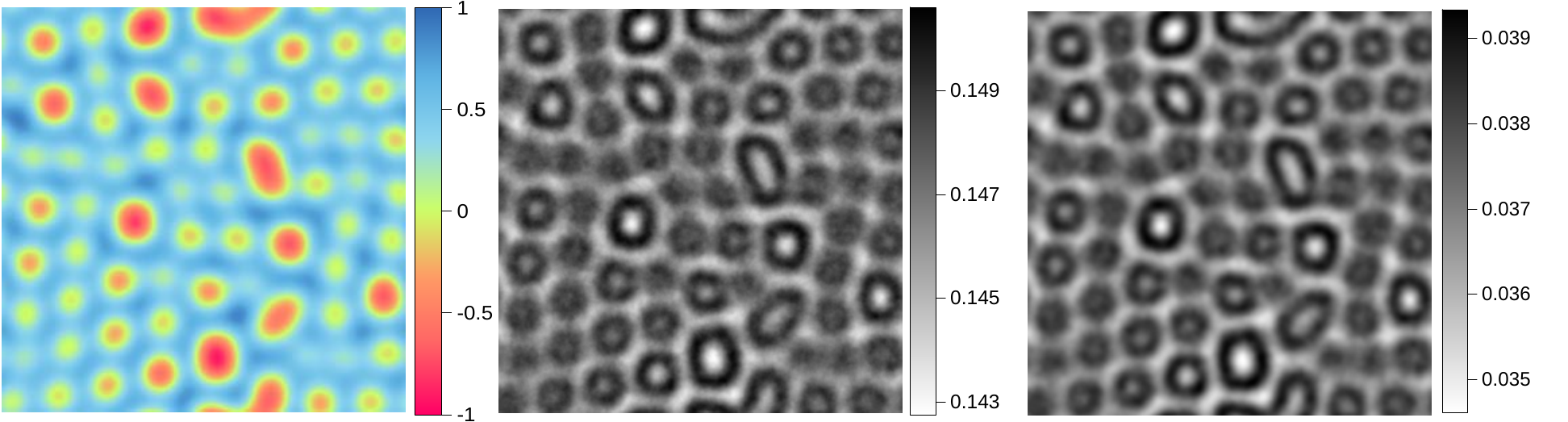}}\\
c)\raisebox{-0.8\height}{ \includegraphics[width=0.78\textwidth]{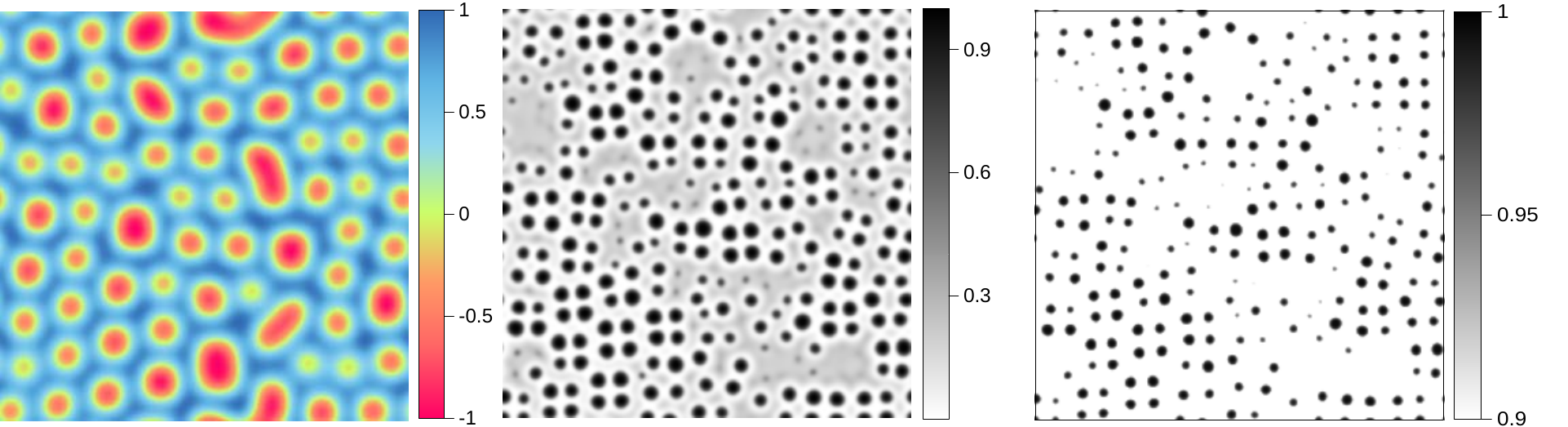}}\\
d)\raisebox{-0.8\height}{ \includegraphics[width=0.78\textwidth]{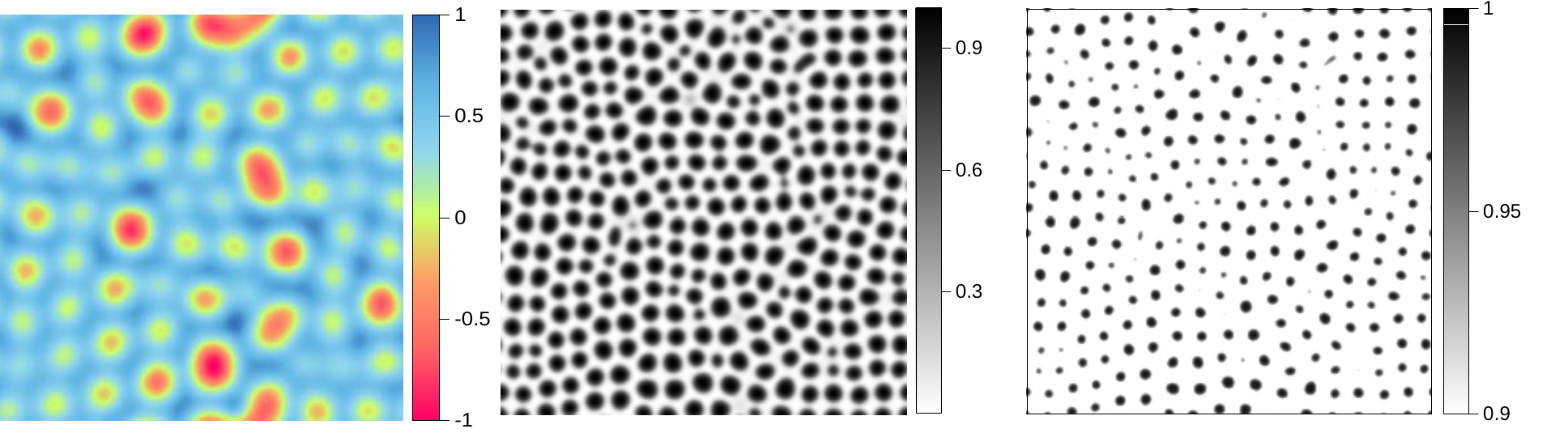}}
\caption{(Colour online)  Snapshots of the system microstructure (solute concentration, vacancy concentration, order parameter, from left to right) at dose $5$~dpa and different irradiation conditions: 
a)~$\mathcal{K}_0\simeq2\times 10^{-3}$~dpa/s, $T\simeq1041$~K;
b)~$\mathcal{K}_0\simeq5\times 10^{-4}$~dpa/s, $T\simeq1041$~K;
c)~$\mathcal{K}_0\simeq3\times 10^{-4}$~dpa/s, $T\simeq833$~K;
d)~$\mathcal{K}_0\simeq2\times 10^{-2}$~dpa/s, $T\simeq1136$~K.
\label{fig_irr_snp}}
\vspace{-3mm}
\end{figure*}

We study the dynamics of the system under different irradiation conditions by considering the averaged vacancy concentration $\langle c_{\text v}\rangle$, phase field $\langle \eta\rangle$  (the solute concentration remains constant, $\psi_0$) and dispersions $\langle(\delta c_{\text v})^2\rangle$, $\langle(\delta \eta)^2\rangle$, $\langle(\delta \psi)^2\rangle$ shown in figure~\ref{fig_irr}. Dispersions of the corresponding values play the role of effective ``order parameters'' for  spatial patterns formation: if these dispersions grow, then spatial  ordering is realized.  Let us start with the simplest case where solid solution is realized (see  dash-dot lines in plots in figure~\ref{fig_irr}). Here,  $\langle c_{\text v}\rangle$ and $\langle \eta\rangle$ grow attaining their stationary values with small values. The corresponding dispersions $\langle(\delta c_{\text v})^2\rangle$, $\langle(\delta \eta)^2\rangle$ tend to zero, which means that no ordering processes are realized in the system. The dispersion $\langle(\delta \psi)^2\rangle$ goes down with the dose growth and takes up a stationary, negligibly small value which means dissolution of the precipitates. Snapshots obtained at dose $5$~dpa are shown in figure~\ref{fig_irr_snp}~(a).
In the case related to a phase decomposition (see dot lines in plots in figure~\ref{fig_irr}), both mean vacancy concentration and phase field attain stationary values ten times larger than in the previous case. By considering the dynamics of the corresponding dispersions, one finds that they initially grow and  attain a small but nonzero values at large doses as well as  the quantity $\langle(\delta\psi)^2\rangle$. It means that here the vacancies produced at irradiation lead to smearing interfaces which causes a slow homogenization  of composition difference  comparing to the previous case (precipitates remain in the system but their interface width is smeared). Here, vacancies mostly segregate at interfaces [see snapshots in  figure~\ref{fig_irr_snp}~(b)], a threshold for vacancy supersaturation causing the formation of voids  is not achieved.
In the domain of pattern formation (see solid and dash lines in plots in figure~\ref{fig_irr}), one observes a more complicated picture of the point defect self-organization. Mean vacancy concentration and phase field initially attain metastable states, and, at a further irradiation,  their values  start to grow until steady states are attained. At this metastable state, accumulation of the defects in vacancy clusters occurs and the voids are  formed when a supersaturation threshold is achieved. The computed critical vacancy supersaturation lies in the interval $1800<c_{\text v}/c_{\text{v}0}<2000$ depending on the irradiation conditions which corresponds well with the previous analysis of the voids formation discussed in \cite{Part1}. Therefore, a growth in $\langle c_{\text v}\rangle$ and $\langle \eta\rangle$ after reaching the metastable state indicates the nucleation and growth of the voids. The slowing-down dynamics at elevated doses corresponds to the voids growth by absorbing the vacancies from the bulk and small voids. Ordering of the system (formation of voids during vacancy patterning) is well seen from protocols of dispersions  $\langle(\delta c_{\text v})^2\rangle$, $\langle(\delta \eta)^2\rangle$. Their growth in the interval $\sim 2{-}2.5$~dpa indicates the formation of the ordered structure of the voids, whereas a further decrease at $\sim 5.2{-}10$~dpa denotes  a growth of large voids by Ostawald ripening. A decrease in the dispersion of the solute concentration means homogenization of the composition difference $\psi$ due to the production of point defects increasing the athermal atomic mixing. A decrease in  $\langle(\delta \psi)^2\rangle$ in this case does not mean homogenization of the alloy. It will be seen below that point defect patterning results in the patterning of alloy components. The corresponding snapshots at $5$~dpa obtained at different irradiation conditions are shown in figure~\ref{fig_irr_snp}~(c),~(d).\footnote{We choose  the level set to distinguish the voids from small vacancy clusters  $\eta_{\text{th}}=0.8$ by taking into account that vacancy relaxation volume is around $0.2\Omega_0$, therefore, without loss of generality, the threshold for $\eta$ related to the voids formation can be formally decreased from 1 down to 0.8 (see discussions in  \cite{Rokkam2009}).}
\begin{figure}[!t]
\centering
a)\raisebox{-0.8\height}{ \includegraphics[width=0.46\textwidth]{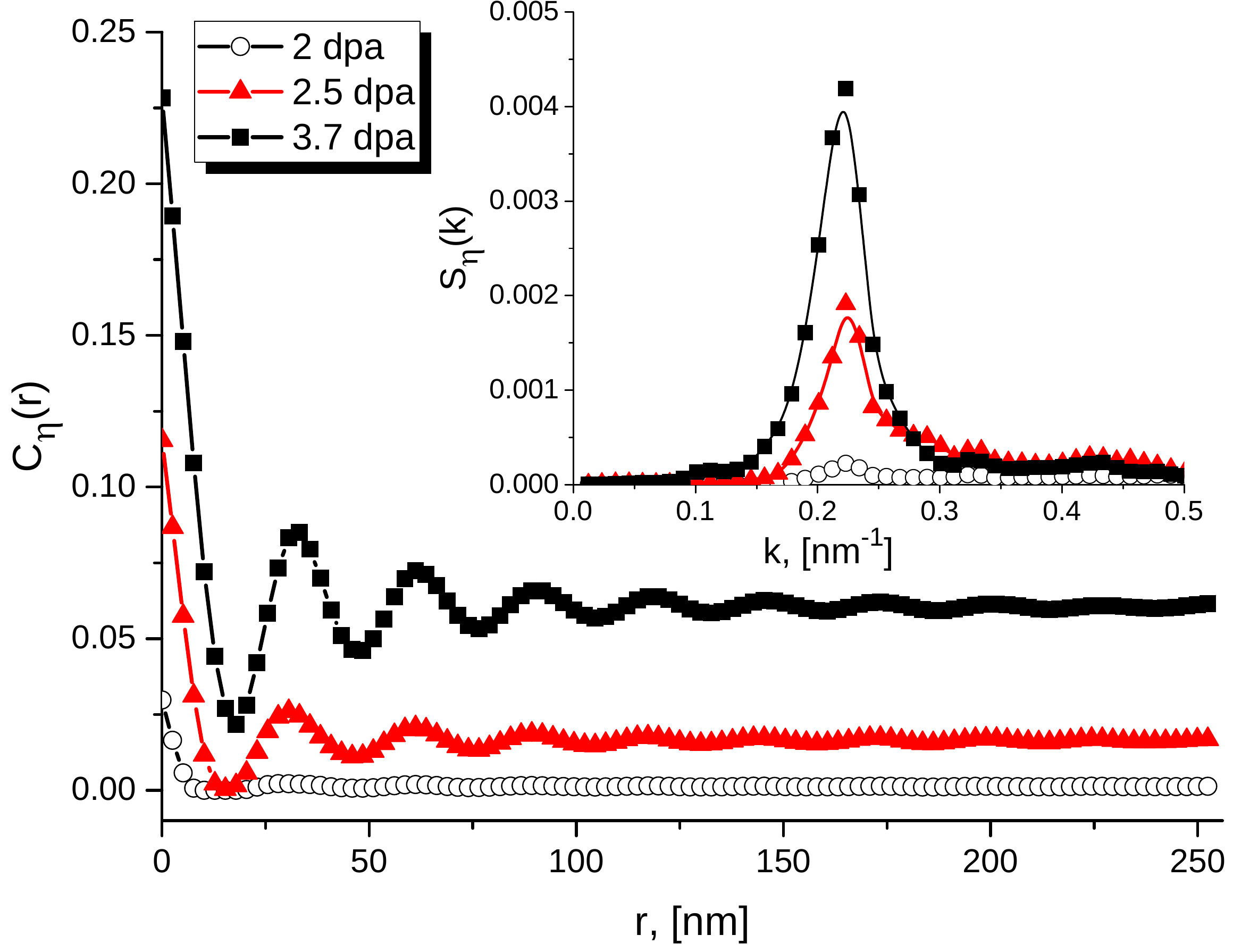}}
b)\raisebox{-0.8\height}{ \includegraphics[width=0.46\textwidth]{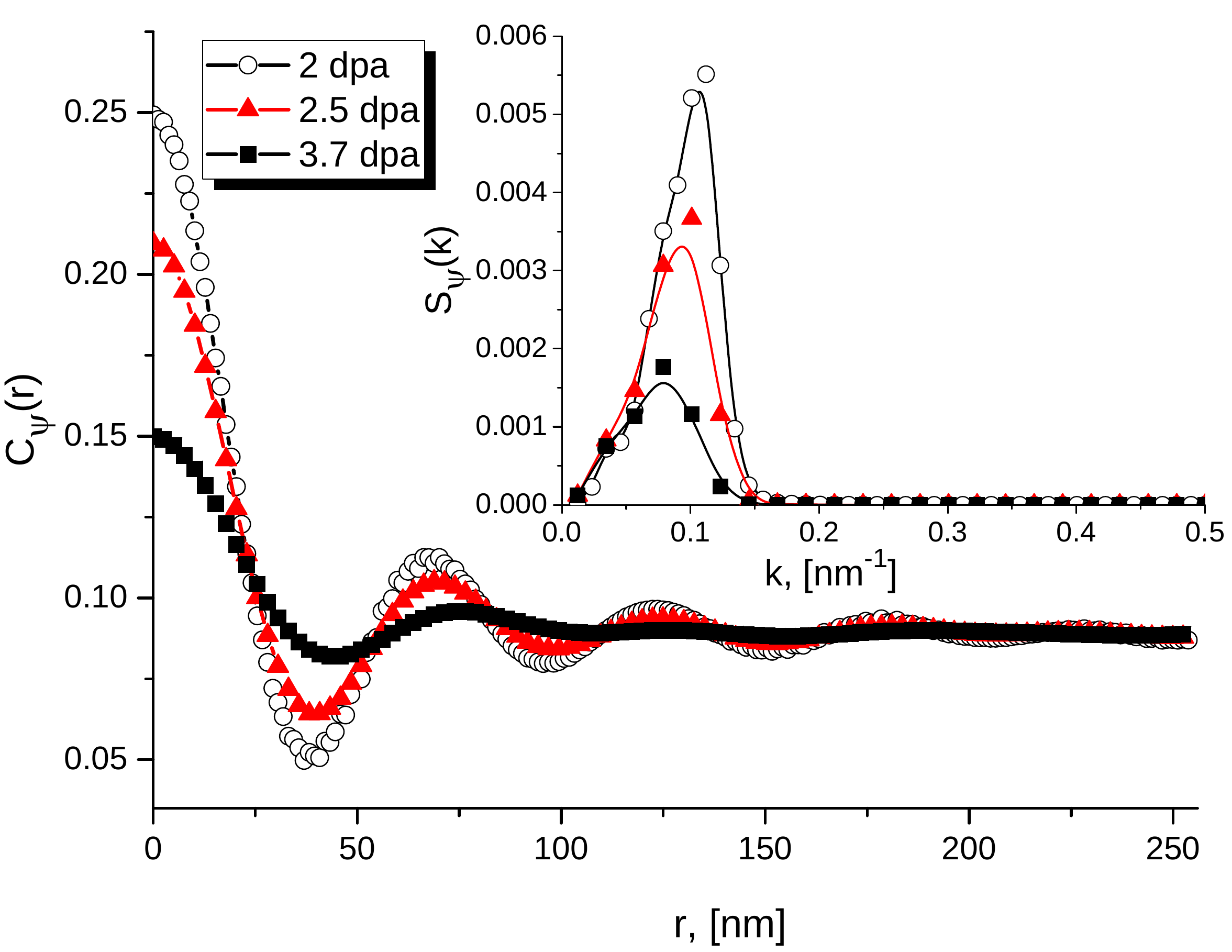}}
\caption{(Colour online)  Dynamics of the spatial correlation functions $G_\eta(r-r',t)$ and  $G_\psi(r-r',t)$ in plots (a) and (b) at $\mathcal{K}_0=10^{-2}$~dpa/s, $T=1041$~K. The corresponding structure functions $S_\eta(k,t)$  $S_\psi(k,t)$  are shown in insertions.
\label{fig_cor_fnc}}
\end{figure}

In the prepared target, one has a quasi-periodic location of precipitates having anisotropic shape formed to reduce the elastic deformation of the lattice. It  leads to the formation  of voids close to the  interfaces having large curvature. Therefore, the voids emergent during irradiation organize into a super-lattice. The corresponding correlation analysis illustrates a periodicity of the voids location. To study this effect, we consider two-point correlation functions: for voids and for precipitates $G_\varrho\psi(r-r', t)=\langle\delta\varrho(\mathbf{r},t)\delta\varrho(\mathbf{r}',t)\rangle$, $\varrho=\{\eta,\psi\}$  with the corresponding Fourier transforms $S_\varrho(k,t)$ shown in figures~\ref{fig_cor_fnc}~(a),~(b), respectively. Well pronounced oscillatory dynamics of $G_\eta(r)$ at the dose growth indicates the formation of statistically periodic void patterns [see figure~\ref{fig_cor_fnc}~(a) and snapshots in figure~\ref{fig_irr_snp}]. The dynamics of the structure function $S_\eta(k,t)$  is shown in the insertion in figure~\ref{fig_cor_fnc}~(a). The position of its peak defines the mean distance between the voids. With the dose growth,  the position of this peak is  slightly shifted toward smaller wave-numbers and its height  increases, meaning the formation of periodically organized voids of a larger size. By studying the dynamics  of both $G_\psi(r-r', t)$ and  $S_\psi(k,t)$, one finds that the oscillatory dynamics becomes less pronounced with the dose accumulation. 
The height of $G_\psi(0)$ decreases  with dose growth  which means that the solute becomes less correlated  (homogenization of the composition difference). The peak of $S_\psi(k,t)$  decreases  and shifts toward small values of the wave number, which corresponds to  homogenization of the field $\psi$. 
An estimation of the void super-lattice parameter shows that the voids are located in their super-lattice at a mean distance around $30{-}40$~nm which is in a good correspondence with the experimental data observed at actual doses and approximately at the same irradiation conditions \cite{GWZ,voids_exp1,voids_exp2}.

\begin{figure}[!t]
\centering
\includegraphics[width=0.495\textwidth]{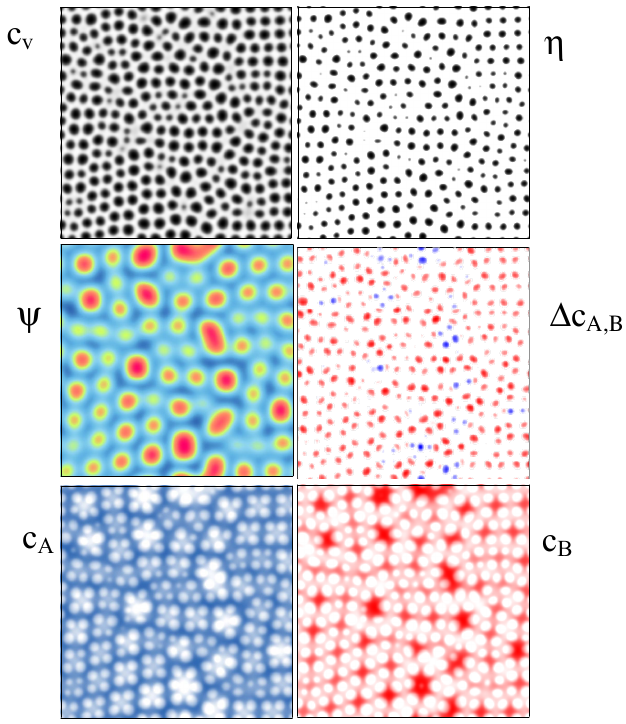}
\caption{(Colour online) Snapshots of distribution of vacancies $c_{\text v}$, phase field $\eta$, composition difference~$\psi$, deviation of concentration of atoms (atoms leaving their positions for the voids formation) $\Delta c_{\text{A,B}}$ (blue and red), and concentration of atoms of A and B sorts at dose $3$~dpa. Other parameters are:  $\mathcal{K}_0\simeq 2\times10^{-2}$~dpa/s, $T\simeq1136$~K. \label{maps}}
\end{figure}

It should be noted that the composition difference $\psi=c_{\text A}-c_{\text B}$ cannot give a complete information about  rearrangement of atoms of each sort. Indeed,  locally, the quantity $\psi$ can be zero in two cases, namely  $c_{\text A}\simeq c_{\text B}\ne 0$ and $c_{\text A}=c_{\text B}\simeq0$. Obviously, the last case corresponds to the voids formation. 
By using the relation  $c_{\text A}+c_{\text B}+c_{\text v}=1$, one can find a schematic distribution of atoms of each sort and obtain a map for concentrations of atoms  moved from their positions by ``liberating positions'' for the voids. The corresponding schematic maps are shown in figure~\ref{maps}.
\begin{figure}[!t]
\centering
a)\raisebox{-0.8\height}{ \includegraphics[width=0.45\textwidth]{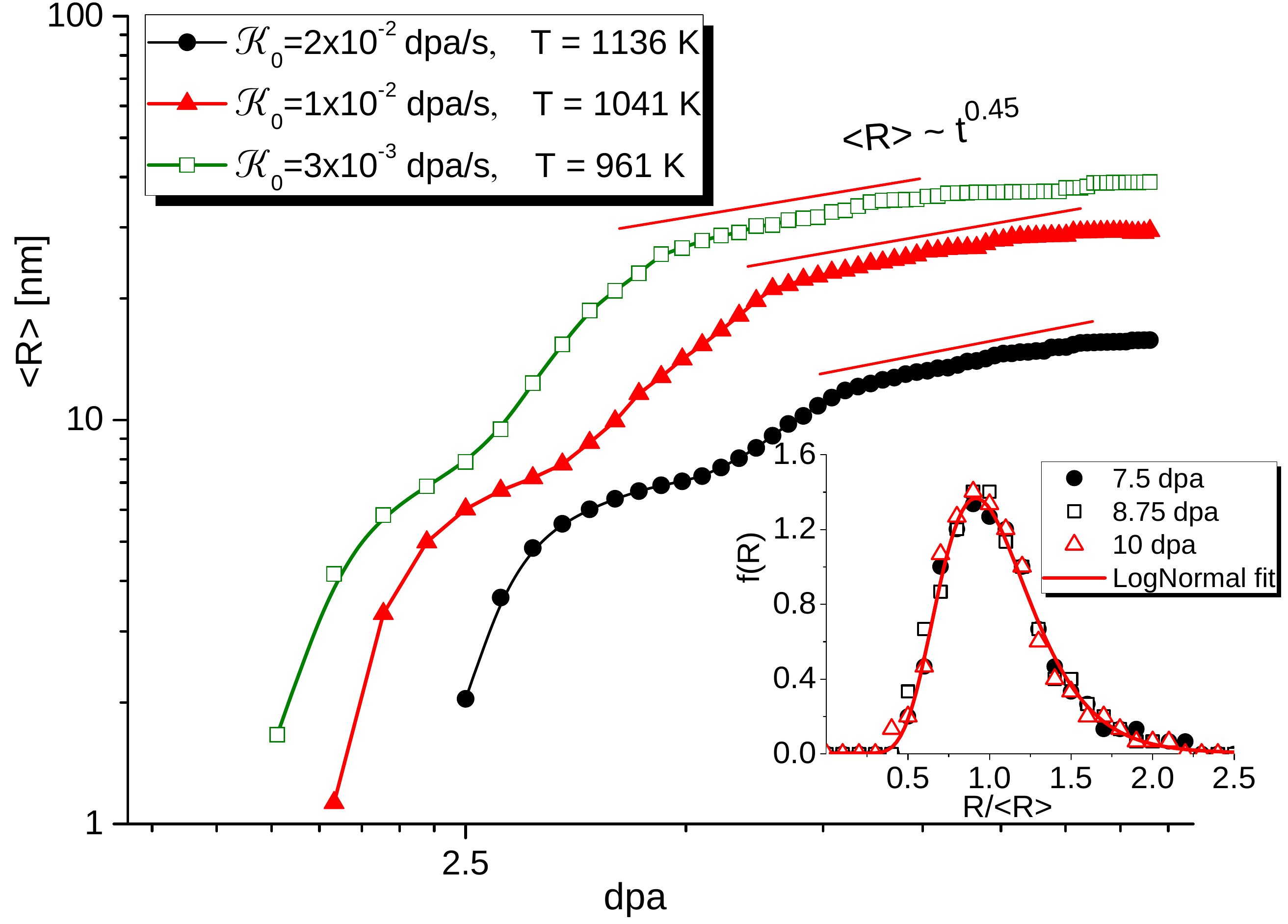}}\
b)\raisebox{-0.8\height}{ \includegraphics[width=0.45\textwidth]{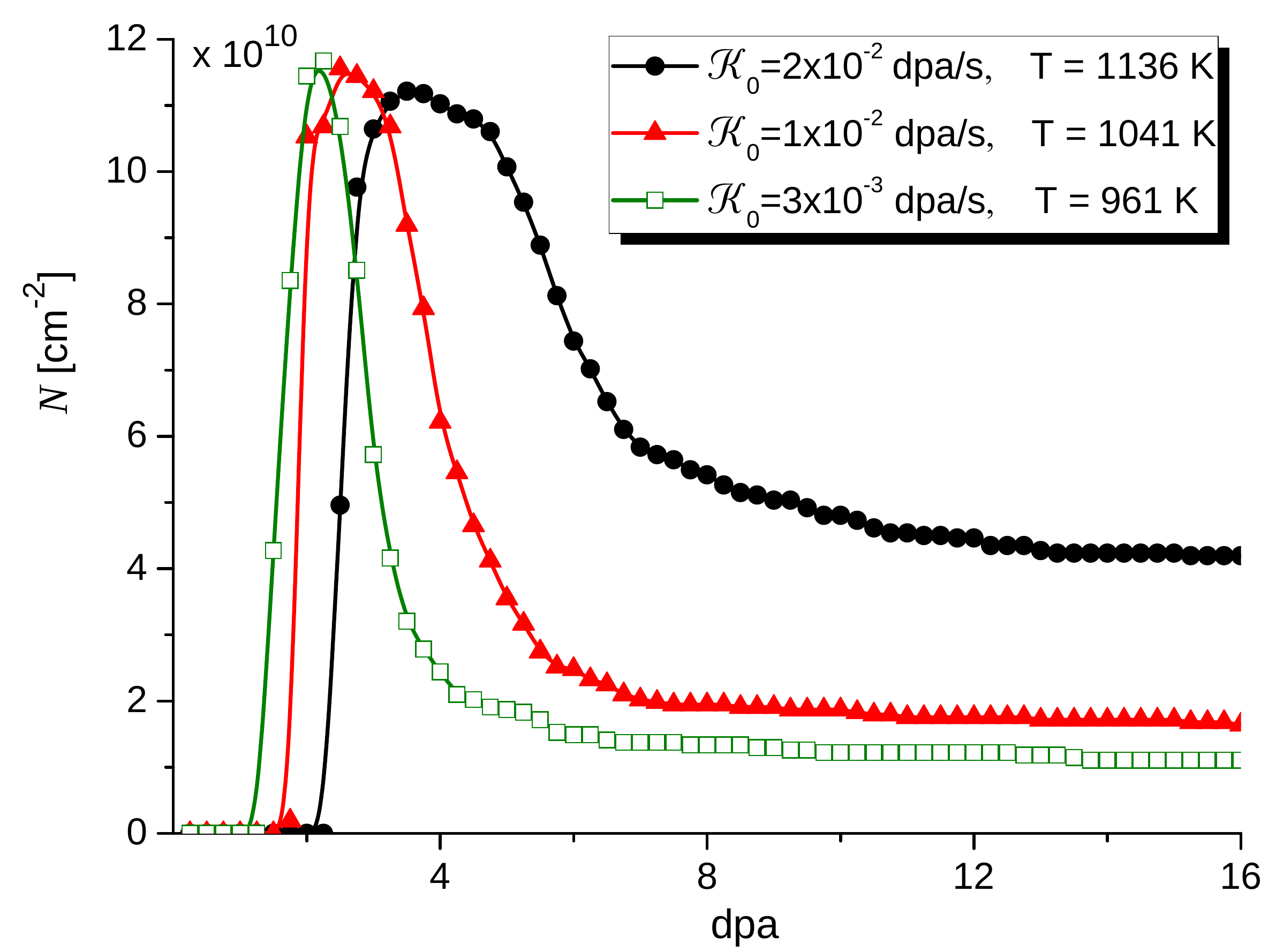}}\\
c)\raisebox{-0.8\height}{ \includegraphics[width=0.45\textwidth]{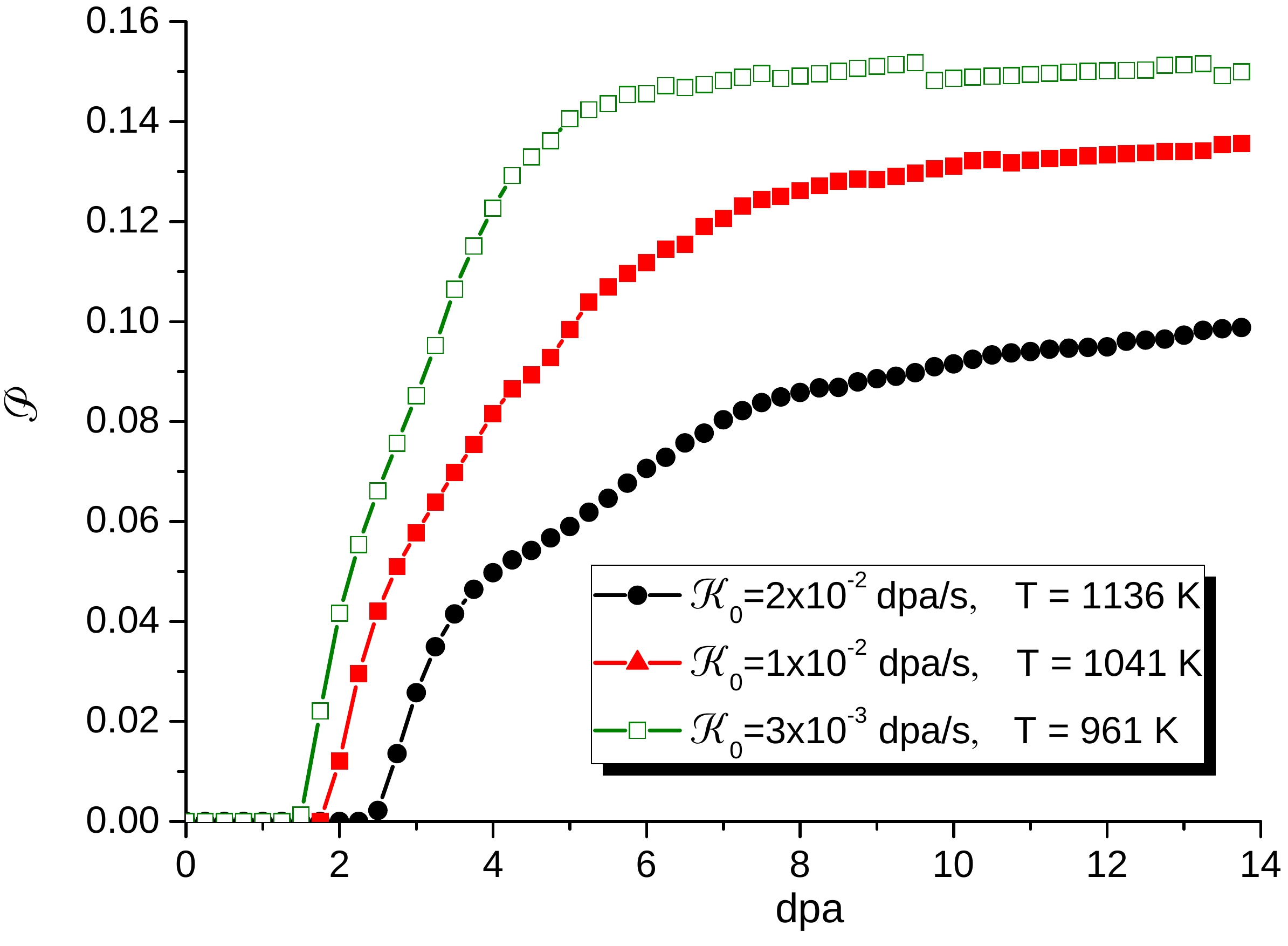}}
\caption{(Colour online) Dynamics of the mean void radius $\langle R\rangle$, voids concentration $\mathcal{N}$  and porosity (void fraction) $\mathcal{P}$ at different irradiation conditions, plots  (a) and (b), and (c), respectively.
\label{fig_irr_RN}}
\end{figure}
It is seen that the voids formation leads to patterning in distribution  of atoms of two sorts resulting in the change of local chemical and, therefore, physical  properties of the alloy. By considering the concentration of atoms leaving their positions for the voids formation  ($\Delta c_{\text{A,B}}$), one finds that most of such atoms belong to  interfaces and soft phase, where vacancy concentration in an unirradiated alloy was large. 

Next, to make a quantitative analysis,  we consider the dynamics of mean void size $\langle R\rangle$, void concentration $\mathcal{N}=n/L^2$, where $n$ is the void number  and porosity $\mathcal{P}$ of the material obtained as the void fraction shown in figure~\ref{fig_irr_RN}.
According to the obtained protocols, one finds that small nuclei emerge after an incubation period.  In our  case, the incubation period is around $1{-}2$~dpa which corresponds well with experimental observations of the voids formation in stainless steels discussed in \cite{GWZ,voids_exp1,voids_exp2}.
In the next stage, these  nuclei grow during irradiation by absorbing the vacancies from the matrix. This stage corresponds to nucleation and growth regime. Here, mean void size increases together with the number of voids. A decrease in the void concentration  $\mathcal{N}$ corresponds to the stage of voids growth by Ostwald ripening mechanism, where large voids grow at the expense of small ones and by absorbing the vacancies from the matrix. At the same time, one can observe the voids coalescence, where closely located voids nucleate and form one large void (see figure~\ref{fig_snap_irr} at $5$~dpa). At this stage, the void mean radius behaves as $\langle R\rangle\propto (\mathcal{K}_0t)^{z}$, where $z$ is the growth exponent. An estimation for this exponent gives $z\simeq0.45$. It means that our approach corresponds well with diffusion controlled precipitation processes described by the  asymptotic  $t^{1/2}$. The emergence of such an asymptotic for the mean void size corresponds with the universal behaviour of the whole system. Here, void distribution and all statistical quantities manifest a power-law scaling.
The void size distribution $f(R)$ is shown in the insertion in figure~\ref{fig_irr_RN}~(a) for the case  $\mathcal{K}_0\simeq2\times 10^{-2}$~dpa/s, $T\simeq1136$~K at different doses. By using a fitting procedure, it was found that the corresponding distribution can be well approximated by the Log-Normal distribution. From the obtained data one can find that the void concentration varies in the interval $10^{10}{-}10^{11}$~cm$^{-2}$. 
The void mean radius varies in the interval $2{-}30$~nm at doses up to $10$~dpa which corresponds well with  experimental data for the voids formation and growth in most of metals and steels \cite{WSSM75,voids_exp1,voids_exp2}. 
\begin{figure}[!b]
\centering
a)\raisebox{-0.8\height}{ \includegraphics[width=0.46\textwidth]{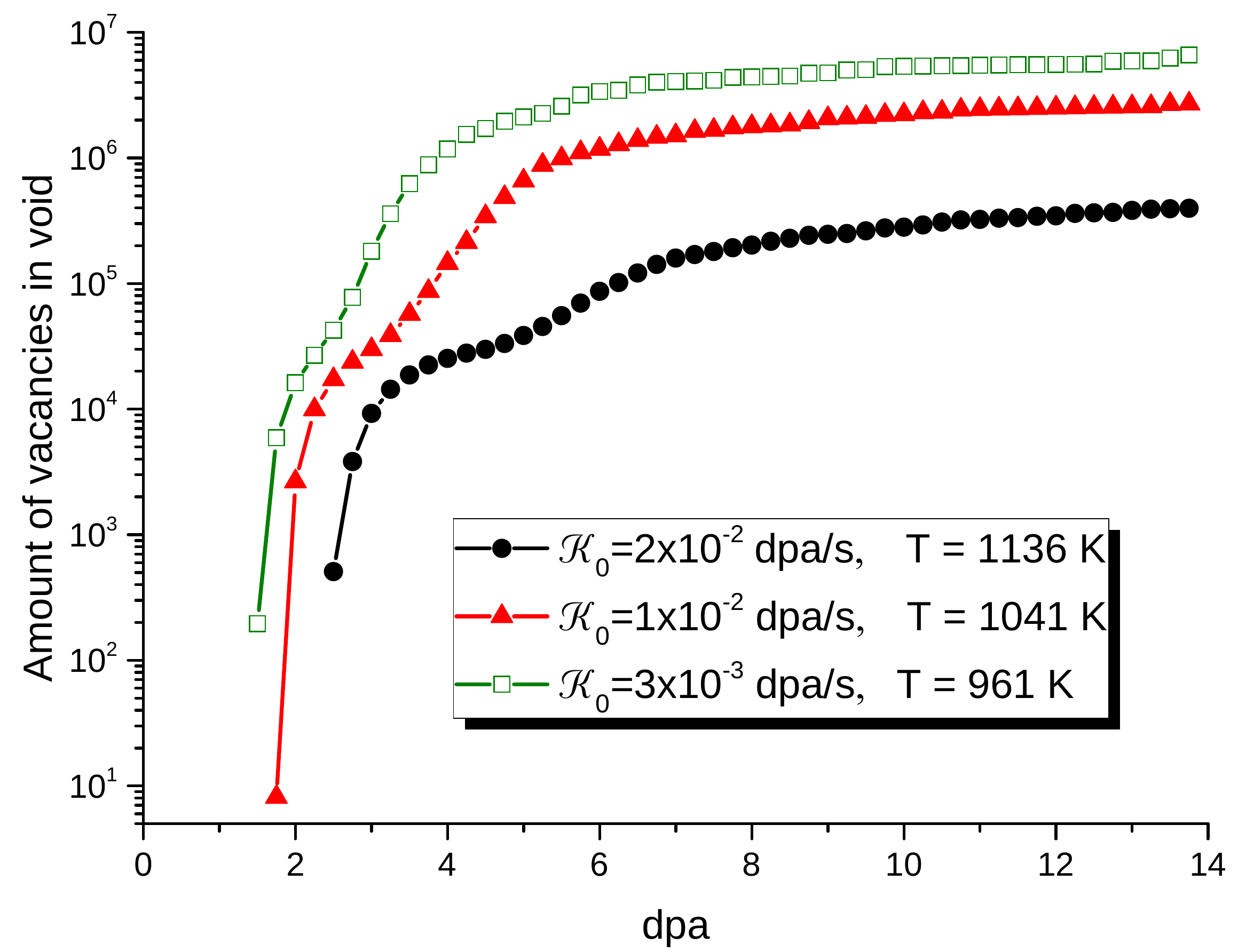}}\
b) \raisebox{-0.8\height}{\includegraphics[width=0.47\textwidth]{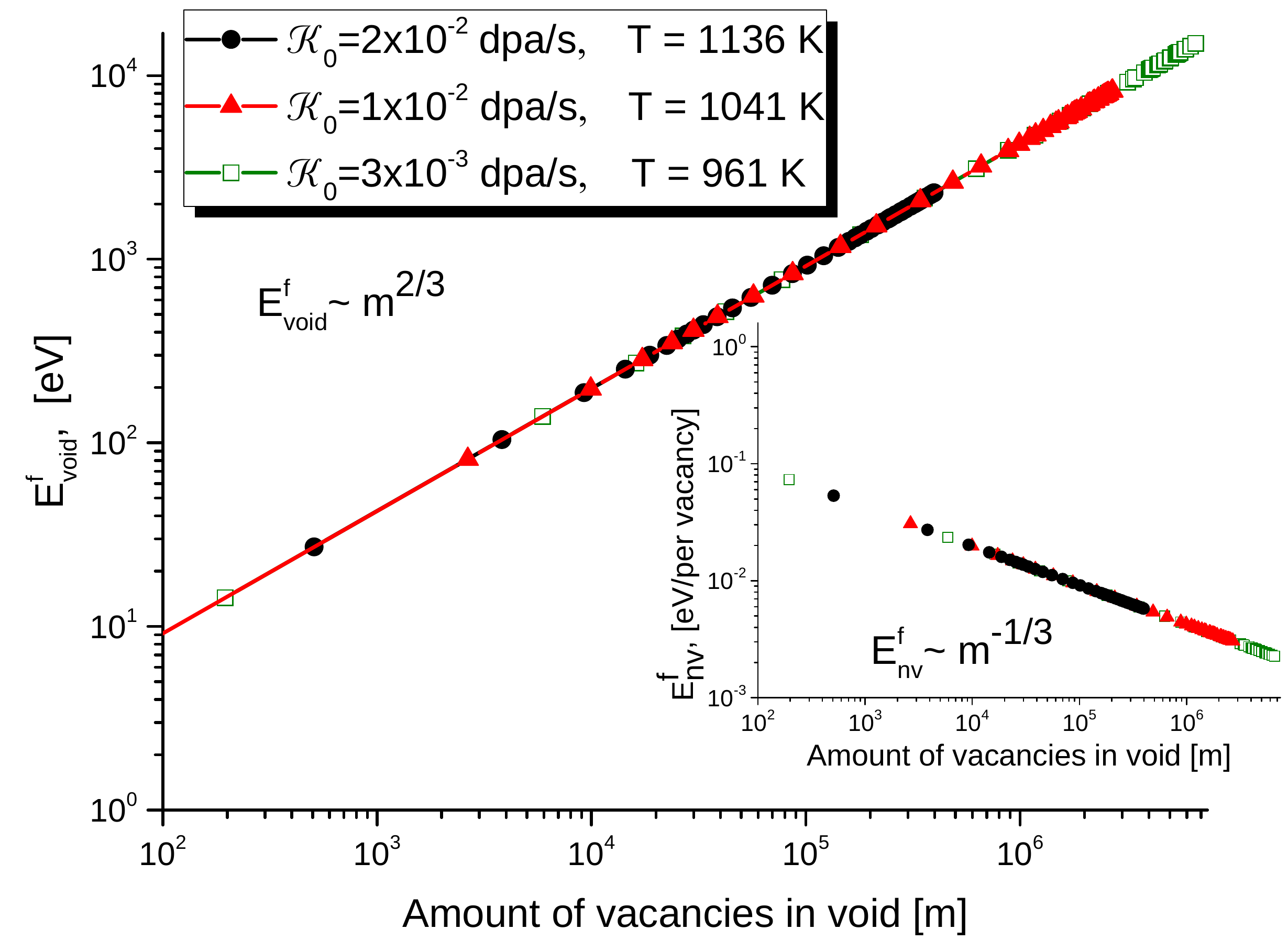}}\\
c) \raisebox{-0.8\height}{\includegraphics[width=0.46\textwidth]{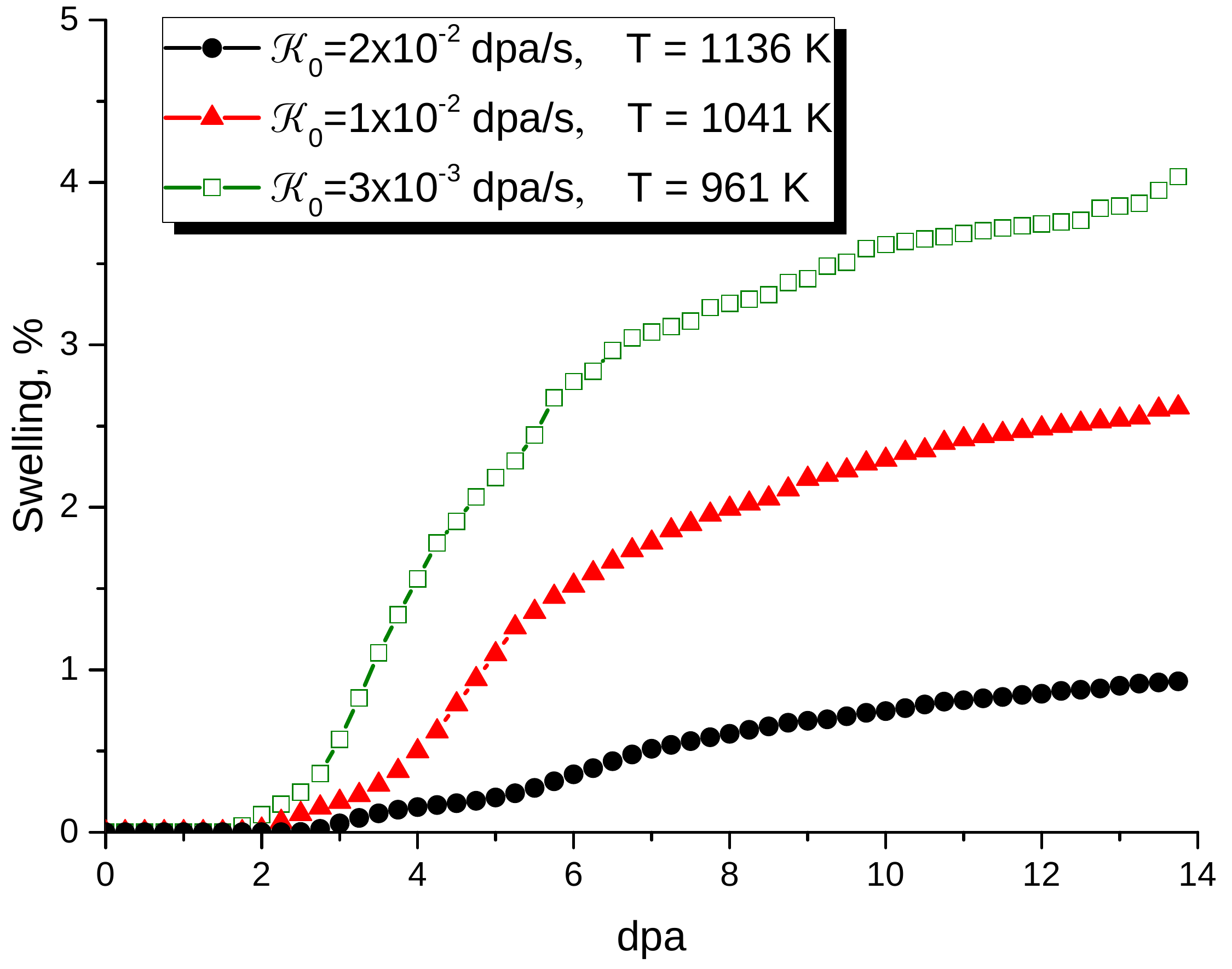}}
\vspace{-1mm}
\caption{(Colour online) (a) The amount of vacancies in the void $m$, (b) void formation energy $E^{\text f}_{\text{void}}$ and voids swelling~(c) at different irradiation conditions.
\label{fig_irr_mES}}
\end{figure}

The obtained data for the voids size can be used to estimate the number of vacancies in growing voids, void swelling, energy of void formation of a fixed size.  
By using  vacancy relaxation volume $\Omega_{\text v}=0.2\Omega_0$, one finds that the number of vacancies in the voids of the linear size $\langle R\rangle$   
is $m\approx4\piup\langle R\rangle^3/3\Omega_{\text v}\propto (\mathcal{K}_0t)^{3z}$. By taking the void surface energy $\gamma_{\text v}\simeq 1.3$~J/m$^2$ for the void formation energy, we use the relation  $E_{\text{void}}^{\text f}=4\piup\langle R\rangle^2\gamma_{\text v}\propto(\mathcal{K}_0t)^{2z}$. The corresponding dependencies of the amount of vacancies in the voids and the void formation energy are shown in figure~\ref{fig_irr_mES}~(a),~(b), respectively.  It is seen that the quantity $m$ grows from   $10^2$ up to $10^6$. The energy needed to form a void of a  given size varies from  $250$~eV up to $3.8$~keV and  behaves as $m^{2/3}$.  The energy needed to put additional vacancies close to the void having $m$ vacancies is $E^{\text f}_{m\text{v}}=E^{\text f}_{\text{void}}/m$. This is shown in the insertion in figure~\ref{fig_irr_mES}~(b). This energy decreases as $m^{-1/3}$, exactly as nucleation theory predicts. It means that  the energy penalty to form a vacancy close to the growing void decreases, and most of the vacancies formed around a large void will be absorbed by the void with a large efficiency $\propto 1/E^{\text f}_{m\text{v}}$.  The computed void swelling is shown in figure~\ref{fig_irr_mES}~(c). It follows that in the actual interval for a dose up to 15~dpa, we get a void swelling up to 5\%. The estimation of the void concentration in a bulk gives $10^{15}{-}10^{16}$~cm$^{-3}$ which corresponds  to experimental  observations in construction materials and steels \cite{WSSM75,voids_exp1,voids_exp2,GWZ}. 

\vspace{-2mm}
\section{Conclusions} \label{5}

In this paper we discussed the dynamics of the voids formation, nucleation and growth  in binary systems with lattice mismatch and elastic inhomogeneity subjected to a constant particle irradiation within the framework of the generalized phase field approach combined with the reaction rate theory. In our study, a structural disorder  is described by the production of point defects  through irradiation in a stochastic manner, where  nonequilibrium point defects lead to an emergence of irradiation-induced mobility of solute atoms resulting in an additional (athermal)  atomic mixing. By using the derived formalism we have studied the point defects rearrangement and patterning of alloy components 
 accompanied by the voids formation, their nucleation and growth.

By using analytical and numerical
analysis, we have computed the phase diagrams and have shown that depending on irradiation conditions  solid solution, phase decomposition, and patterning are observed. The  solid solution is characterized by dissolved precipitates of an initially prepared target with a homogeneous vacancy distribution;  phase decomposition relates to segregation of vacancies at interfaces and in the soft phase; patterning corresponds to the voids formation. The obtained results qualitatively relate  to the known theoretical ones concerning the problems of phase decomposition and patterning of binary systems under irradiation \cite{Martin,Vaks,BBA2015,EB2000,EB2001,Nastar}.

The main attention in this article was paid to the study of the processes of the voids formation and their growth. We have shown that elastic inhomogeneity in the target leading to the formation of anisotropic precipitates of the soft phase results in the formation of a super-lattice of the voids located at phase interfaces having a large curvature. It was shown that the voids nucleate and grow with the dose according  to diffusion controlled precipitation processes by the Ostwald ripening mechanism. We have found that the growth of the voids is related to patterning of alloy components. By using statistical analysis, we have shown that the voids growth dynamics is universal with the growth exponent being around the theoretically predicted value 1/2. At the same time, it is found that the void size distribution is of Log-Normal form independent of the irradiation conditions (temperature and damage rate). 

By using material parameters relevant to most of alloys and steels, we have estimated the main statistical characteristics of the voids. We have found that critical vacancy supersaturation leading to the voids formation depends on irradiation conditions and is of three orders of equilibrium vacancy concentration.
It is shown that  at actual range of irradiation temperature and dose rate, the voids  size  varies in the interval from 2 up to $30$~nm with a super-lattice parameter  of the size $30{-}40$~nm with the dose growth. 
The estimated number of vacancies in the voids varies in the interval $10^2$ up to $10^6$, whereas the voids concentration in a bulk is of the order  $10^{15}{-}10^{16}$~cm$^{-3}$. Computed  void swelling at doses up to $10$~dpa does not exceed value of 5\%. The quantitative results obtained within this work  correspond well to  most of experimental observations of the voids formation in the alloys and steels irradiated by ions with the actual range of dose rate and temperature \cite{WSSM75,voids_exp1,voids_exp2,GWZ,Was,voidsNiAl}. 

The obtained results can be useful to predict the development of a defect structure and the voids formation in the alloys at different irradiation conditions.
The derived model can be generalized to study the mechanical properties of irradiated alloys and the alloys under sustained irradiation at different external loadings.

We expect that our nontrivial findings will stimulate further theoretical and experimental studies of the voids formation processes in the alloys under extreme nonequilibrium conditions. 

\vspace{-2mm}
\section*{Acknowledgements}
This work was financially supported by National Nature Science Foundation of China (51601185),  Sichuan Province International Science and
Technology Cooperation and Exchange Research Program (2016HH0014) and also the
China Postdoctoral Science Foundation (2015M582575).

\ukrainianpart

\title{Фазово-польове моделювання нуклеації та росту пор в бінарних системах}
\author{Д.О. Харченко\refaddr{label1}, В.О. Харченко\refaddr{label1}, Ю.М. Овчаренко\refaddr{label1}, О.Б. Лисенко\refaddr{label1}, І.О. Шуда\refaddr{label2}, В.~Лу\refaddr{label3}, Р. Пан\refaddr{label3} }
\addresses{
\addr{label1} Інститут прикладної фізики НАН України,  вул. Петропавлівська, 58, 40000 Суми, Україна
\addr{label2} Сумський державний університет, вул. Римського-Корсакова, 2, 40007 Суми, Україна 
\addr{label3} Перший інститут, Інститут ядерної енергії Китаю,  Перша секція, дор. Чангшундадао, 328,  Шуангліу,  610213 Ченгду, Китай
}

\makeukrtitle

\begin{abstract}
\tolerance=3000%
Проводиться всебічне дослідження процесів формування та росту пор у типовій моделі бінарного стопу, підданому радіаційному впливові з використанням комбінованого підходу, що грунтується на теорії фазового поля та щвидкісній теорії. Показано, що формування пор викликане взаємодією радіаційно продукованих вакансій з пружніми деформаціями гратниці та полем композиції стопу. Отримано фазові діаграми, що ілюструють формування твердого розчину, фазового розшарування та структуроутворення. Встановлено, що формування пор із пересиченого розчину вакансій супроводжується перерозподілом композиційного поля. Показано, що пружна неоднорідність, яка відповідає за утворення анізотропних преципітатів у первинно приготовленому стопі  приводить до формування над\cyrgup ратки пор при опроміненні.
Виявлено, що нуклеація та ріст пор з ростом дози опромінення відповідає дифузійно керованим процесам випадіння фаз. Показано універсальність динаміки росту пор. Проведена оцінка основних параметрів та статистичних характеристик пор для більшості стопів та сталей дає добре узгодження з експериментально отриманими результатами.

\keywords бінарні стопи, структуроутворення, опромінення, точкові дефекти, пори
\end{abstract}


\begin{thebibliography}{99}

\bibitem{ZS1999}Zinkle S.J., Singh B.N., J. Nucl. Mater., 1993, \textbf{199}, 173--191,
\doi{10.1016/0022-3115(93)90140-T}.

\bibitem{clusters1}Hulett L.D. (Jr.), Baldwin T.O., Crump J.C. III, Young F.W. (Jr.), J. Appl. Phys., 1968, \textbf{39}, 3945--3954, \doi{10.1063/1.1656879}. 

\bibitem{clusters2}Brimhall J.L., Kulcinski G.L., Kissinger H.E., Mastel B., Radiat. Eff., 1971, \textbf{9}, 273--278,\\ \doi{10.1080/00337577108231060}.
 
\bibitem{walls}Stiegler J.O., Farrell K., Scr. Metall., 1974, \textbf{8}, 651--655, 
\doi{10.1016/0036-9748(74)90015-5}.

\bibitem{DM2003}Doan N.V., Martin G., Phys. Rev. B, 2003, \textbf{67}, 134107, \doi{10.1103/PhysRevB.67.134107}.

\bibitem{Walgraef1} Walgraef D., Spatio-Temporal Pattern Formation: With Examples from Physics, Chemistry, and Material Science, Springer-Verlag, New York, 1997, \doi{10.1007/978-1-4612-1850-0}.

\bibitem{Was} Was  G.S.,  Fundamentals of Radiation Materials Science: Metals and Alloys, Springer, Berlin, Heidelberg,  2007, \doi{10.1007/978-3-540-49472-0}.

\bibitem{Evans}  Evans J.H., van Veen A.,   Caspers L.M., Nature, 1981, \textbf{291}, 310--312, \doi{10.1038/291310a0}.

\bibitem{voidsAl} Horsewell A., Singh B.N., Radiat. Eff., 1987, \textbf{102}, 1--5,
\doi{10.1080/00337578708222901}.

\bibitem{voidsNb} Brimhall J.L., Kissinger H.E., Kulcinski G.L., In: Radiation-Induced Voids in Metals: Proceedings, Corbett~J.W., Ianniello~L.C.~(Eds.), National Technical Information Service, Springfield, VA, 1972,  338--362.  

\bibitem{voidsNi} Kulcinski G.L., Brimhall J.L., Effects of Radiation on Substructure
and Mechanical Properties of Metals and Alloys, American Society for Testing and Materials, Philadelphia,  1973.

\bibitem{voidsNiAl} Chen L.J., Ardell A.J.,  J. Nucl. Mater., 1978, \textbf{75}, 177--185, 
\doi{10.1016/0022-3115(78)90042-9}.

\bibitem{voidsCuNi}Zinkle S.J., Singh B.N.,  J. Nucl. Mater., 2000, \textbf{283--287}, 306--312,
\doi{10.1016/S0022-3115(00)00359-7}.

\bibitem{voidsNbZr} Loomis B.A., Gerber S.B.,  Taylor A.,  J. Nucl. Mater., 1977, \textbf{68}, 19--31, 
\doi{10.1016/0022-3115(77)90212-4}.

\bibitem{voidsMoTi} Evans J.H., In: Proceedings of the International Conference on Irradiation Behavior of Metallic Materials for Fast Reactor Core Components (Ajaccio,  1979), Poirier~J., Dupouy~J.M.~(Eds.), CEA, Gif-Sur-Yvette, 1979, 225. 

\bibitem{voidsSS}Farrell K., Packan N.H.,  J. Nucl. Mater., 1979, \textbf{85--86}, 683--687, 
\doi{10.1016/0022-3115(79)90339-8}.

\bibitem{voids_exp1}Mazey D.J., Hudson J.A., Nelson R.S., J. Nucl. Mater., 1971, \textbf{41}, 257--273, \doi{10.1016/0022-3115(71)90164-4}.

\bibitem{GWZ}Ghoniem N.M., Walgraef D., Zinkle S.J., J. Comput.-Aided Mater. Des., 2001, \textbf{8}, 1--38,\\ \doi{10.1023/A:1015062218246}. 

\bibitem{R1971} Russell K.C., Acta Metall., 1971, \textbf{19}, 753--758,  
\doi{10.1016/0001-6160(71)90131-3}.

\bibitem{W1972}  Wiedersich H., Radiat. Eff., 1972, \textbf{12},  111--125,
\doi{10.1080/00337577208231128}.

\bibitem{BBH1976} Brailsford A.D., Bullough R., Hayns M.R., J. Nucl. Mater., 1976, \textbf{60}, 246--256,\\  \doi{10.1016/0022-3115(76)90139-2}.

\bibitem{MB1980}Mayer R.M.,  Brown L.M., J. Nucl. Mater., 1980, \textbf{95}, 58--63,
\doi{10.1016/0022-3115(80)90080-X}.

\bibitem{WS1982}Wolfer W.G., Si-ahmed A., Philos. Mag. A, 1982, \textbf{46}, 723--736,
\doi{10.1080/01418618208236927}.

\bibitem{GW1993}Ghoniem N.M., Walgraef D., Modell. Simul. Mater. Sci. Eng., 1993, \textbf{1}, 569--590,\\ \doi{10.1088/0965-0393/1/5/001}.

\bibitem{Golubov} Golubov S.I.,  Barashev A.V., Stoller R.E., In: Reference Module in
Materials Science and Materials Engineering, Hashmi~S.~(Ed.), Elsevier, Oxford, 2016, 1--41. 

\bibitem{Walgraef96}Walgraef D., Lauzeral J., Ghoniem N.M., Phys. Rev. B, 1996, \textbf{53}, 14782--14794,\\ \doi{10.1103/PhysRevB.53.14782}.


\bibitem{EPJB2012}Kharchenko V.O., Kharchenko D.O.,  Eur. Phys. J. B, 2012, \textbf{85}, 383,  \doi{10.1140/epjb/e2012-30522-3}.

\bibitem{CMPh2013}Kharchenko V.O., Kharchenko D.O., Condens. Matter Phys., 2013, \textbf{16}, 33001, \doi{10.5488/CMP.16.33001}. 

\bibitem{UJP2013}Kharchenko D.O., Kharchenko V.O., Bashtova A.I., Ukr. J. Phys., 2014, \textbf{58}, No. 10, 993--1008,\\ \doi{10.15407/ujpe58.10.0993}. 

\bibitem{REDS2014}Kharchenko D.O., Kharchenko V.O., Bashtova A.I., Radiat. Eff. Defects Solids, 2014,  \textbf{169}, 418--436,  \doi{10.1080/10420150.2014.905577}. 

\bibitem{EPJB2016} Kharchenko D.O., Kharchenko V.O., Bashtova A.I., Eur. Phys. J. B, 2016, \textbf{89}, 123,\\ \doi{10.1140/epjb/e2016-70090-x}.

\bibitem{PRE2014}Kharchenko V.O., Kharchenko D.O.,  Phys. Rev. E, 2014, \textbf{89}, 042133, \doi{10.1103/PhysRevE.89.042133}. 

\bibitem{Cahn61} Cahn J.W., Acta Metall., 1961, \textbf{9}, 795--801, 
\doi{10.1016/0001-6160(61)90182-1}.

\bibitem{Cahn63}Gahn J.W., Acta Metall., 1963, \textbf{11}, 1275--1282, 
\doi{10.1016/0001-6160(63)90022-1}.

\bibitem{Cahn58} Cahn J.W., Hilliard J.E., J. Chem. Phys., 1958, \textbf{28}, 258--267,
\doi{10.1063/1.1744102}.

\bibitem{LiuBellon2002}Liu J.-W., Bellon P., Phys. Rev. B, 2002, \textbf{66}, 020303(R), 
\doi{10.1103/PhysRevB.66.020303}. 

\bibitem{EB2001} Enrique R.A., Bellon P., Phys. Rev. B, 2001, \textbf{63}, 134111, 
\doi{10.1103/PhysRevB.63.134111}.

\bibitem{EB2000}Enrique R.A., Bellon P., Phys. Rev. Lett., 2000, \textbf{84}, 2885--2888, 
\doi{10.1103/PhysRevLett.84.2885}.

\bibitem{Rokkam2009} Rokkam S., El-Azab A., Millett P., Wolf D., Modell. Simul. Mater. Sci. Eng., 2009, \textbf{17}, 064002, 
\doi{10.1088/0965-0393/17/6/064002}. 

\bibitem{YuLu2005} Yu H.-C., Lu W., Acta Mater., 2005, \textbf{53}, 1799--1807, \doi{10.1016/j.actamat.2004.12.029}. 

\bibitem{VoidsReview2017} Li Y., Hu S., Sun X., Stan M., npj Comput. Mater., 2017, \textbf{3}, 16, \doi{10.1038/s41524-017-0018-y}. 

\bibitem{REDS2015} Kharchenko D.O., Schokotova O.M., Lysenko I.O., Kharchenko V.O., Radiat. Eff. Defects Solids, 2015, \textbf{170}, 584--600, \doi{10.1080/10420150.2015.1063058}. 

\bibitem{Part1} Millett P.C., El-Azab A., Rokkam S., Tonks M., Wolf D., Comput. Mater. Sci.,  2011, \textbf{50}, 949--959,  \doi{10.1016/j.commatsci.2010.10.034}. 

\bibitem{Part2} Millett P.C., El-Azab A., Wolf D., Comput. Mater. Sci., 2011, \textbf{50},  960--970,\\ \doi{10.1016/j.commatsci.2010.10.032}. 

\bibitem{HH2009} Hu S., Henager C.H. (Jr.), J. Nucl. Mater., 2009, \textbf{394}, 155--159, 
\doi{10.1016/j.jnucmat.2009.09.002}.

\bibitem{BBA2015}Badillo A., Bellon P., Averback R.S., Modell. Simul. Mater. Sci. Eng., 2015, \textbf{23}, 035008,\\ \doi{10.1088/0965-0393/23/3/035008}.

\bibitem{Gusak}Gusak A.M., Kornienko S.V., Lutsenko G.V., Defect Diffus. Forum, 2007, \textbf{264}, 109--116,\\ \doi{10.4028/www.scientific.net/DDF.264.109}. 

\bibitem{UJP2016}Kharchenko D.O., Kharchenko V.O., Bashtova A.I., Ukr. J. Phys., 2016, \textbf{61}, No. 3, 265--278,\\ \doi{10.15407/ujpe61.03.0265}. 

\bibitem{REDS2016} Kharchenko D.O., Kharchenko V.O., Radiat. Eff. Defects Solids, 2016, \textbf{171}, 819--839, \\
\doi{10.1080/10420150.2016.1274753}.

\bibitem{Martin} Martin G., Phys. Rev. B, 1984, \textbf{30}, 1424--1436,
\doi{10.1103/PhysRevB.30.1424}.

\bibitem{Vaks} Vaks V.G., Kamyshenko V.V., Phys. Lett. A, 1993, \textbf{177}, 269--274,
\doi{10.1016/0375-9601(93)90039-3}.

\bibitem{BEK} Kulchinski G.L., Brimhall J.L.,  American Society for
Testing and Materials Report No.~ASTM-STP~529, ASTM, Philadelphia, 1973, 258. 

\bibitem{SemWoo}Semenov A.A., Woo C.H., J. Phys. D: Appl. Phys., 2001, \textbf{34}, 3500,
\doi{10.1088/0022-3727/34/24/313}.

\bibitem{ERS} Ehrhart P., Robrock K.H.,  Schober H.R., In: Basic Defects
in Metals, Physics of Radiation Effects in Crystals,  Johnson R.A., Orlov A.N.~(Eds.), Elsevier, Amsterdam, 1986, 3. 

\bibitem{OBSSG}Osetsky Yu.N., Bacon D.J., Serra A., Singh B.N., Golubov S.I.,
 J. Nucl. Mater., 2000, \textbf{276}, 65--77,
\doi{10.1016/S0022-3115(99)00170-1}.

\bibitem{OnukiBook2002}Onuki A., Phase Transition Dynamics, Cambridge University
Press, Cambridge, 2002.

\bibitem{OnukiPhysRevE68}Onuki A., Phys. Rev. E, 2003, \textbf{68}, 061502, 
\doi{10.1103/PhysRevE.68.061502}.

\bibitem{OnukiPhysRevB70} Minami A., Onuki A., Phys. Rev. B, 2004, \textbf{70}, 184114, 
\doi{10.1103/PhysRevB.70.184114}.

\bibitem{OnukiPhysRevB72}Minami A., Onuki A., Phys. Rev. B, 2005, \textbf{72}, 100101(R),
\doi{10.1103/PhysRevB.72.100101}.

\bibitem{Khachaturyan} Khachaturyan A.G., Theory of Structural Transformations in Solids, Wiley, New York, 1983.

\bibitem{Mirzoev}Mirzoev F.Kh., Panchenko V.Ya., Shelepin L.A., Phys. Usp., 1996, \textbf{39}, 1--29,\\ \doi{10.1070/PU1996v039n01ABEH000125}.

\bibitem{OnukiJPhysSocJpn60} Nishimori H.,  Onuki A., J. Phys. Soc. Jpn., 1991, \textbf{60}, 1208--1211, \doi{10.1143/JPSJ.60.1208}.

\bibitem{OnukiPhysRevLet86}Onuki A., Furukawa A., Phys. Rev. Lett., 2001, \textbf{86}, 452--455,
\doi{10.1103/PhysRevLett.86.452}.

\bibitem{OnukiPhysRevB43} Onuki A., Nishimori H., Phys. Rev. B, 1991, \textbf{43}, 13649,
\doi{10.1103/PhysRevB.43.13649}.

\bibitem{DLPS} Demange G., Lun\'eville L., Pontikis V., Simeone D., J. Appl. Phys., 2017, \textbf{121}, 125108, \doi{10.1063/1.4978964}.


\bibitem{MTMA96}Matsumura S., Tanaka Y., M\"uller S., Abromeit C., J. Nucl. Mater., 1996, \textbf{239}, 42--49,\\ \doi{10.1016/S0022-3115(96)00431-X}.

\bibitem{Nastar} Schuler T., Nastar M., Soisson F., Phys. Rev. B, 2017, \textbf{95}, 014113, \doi{10.1103/PhysRevB.95.014113}. 

\bibitem{phd_WA97} Wei L.C., Averback R.S., J. Appl. Phys., 1997, \textbf{81}, 613--623,
\doi{10.1063/1.364202}.

\bibitem{PhysA2017}Kharchenko D.O., Kharchenko V.O., Lysenko I.O., Shuda I.A., Physica A, 2017, \textbf{486}, 497--507,\\ \doi{10.1016/j.physa.2017.05.053}.

\bibitem{Ostwald0}Ostwald W., Z. Phys. Chem., 1901, \textbf{37U},  385. 

\bibitem{Ostwald1}Baldan A., J. Mater. Sci., 2002, \textbf{37}, 2171--2202, 
\doi{10.1023/A:1015388912729}. 

\bibitem{Ostwald} Balluffi R.W., Allen S.M., Carter W.C., Kinetic of Materials, Wiley, New Jersey,  2005, \doi{10.1002/0471749311}.

\newpage

\bibitem{Anisvoids2016} Liu W.B., Wang N., Ji Y.Z., Song P.C., Zhang C., Yang Z.G., Chen L.Q., J. Nucl. Mater., 2016, \textbf{479}, 316--322, \doi{10.1016/j.jnucmat.2016.07.010}. 

\bibitem{Chen73} Chen C.W., Phys. Status Solidi A, 1973, \textbf{16}, 197--210, 
\doi{10.1002/pssa.2210160121}.

\bibitem{voids_exp2}Hudson J.A., Mazey D.J.,  Nelson R.S., J. Nucl. Mater., 1971, \textbf{41}, 241--256, \doi{10.1016/0022-3115(71)90163-2}.

\bibitem{WSSM75} Westmoreland J.E., Sprague J.A., Smidt F.A. (Jr.),
Malmberg P.R., Radiat. Eff., 1975, \textbf{26}, 1--16,\\ \doi{10.1080/00337577508237413}.


\end{thebibliography}
\end{document}